\documentclass[preliminary,copyright,creativecommons]{eptcs}
\providecommand{\event}{MARS 2020} 
\usepackage{xspace}
\usepackage{xcolor}
\usepackage{graphicx}
\usepackage{amsfonts,amsmath,amssymb}
\usepackage{enumitem}
\usepackage{amsmath}
\usepackage{awn,awn_keywords}
\usepackage{subcaption}
\usepackage{uppaal}
\usepackage{xspace}
\setitemize{noitemsep,topsep=0pt,parsep=0pt,partopsep=0pt,leftmargin=0.7cm}
\setenumerate{noitemsep,topsep=0pt,parsep=0pt,partopsep=0pt,leftmargin=0.7cm}

\definecolor{backgroundColour}{rgb}{0.95,0.95,0.92}

\usepackage{hyperref}

\newcommand{\hellomsg}{{\scshape Hello} message\xspace}
\newcommand{\hellomsgs}{{\hellomsg}s\xspace}
\newcommand{\dbdmsg}{{\scshape Dbd} message\xspace}
\newcommand{\dbdmsgs}{{\dbdmsg}s\xspace}
\newcommand{\lsadv}{{\scshape Lsa}\xspace}
\newcommand{\lsadvs}{{\lsadv}s\xspace}
\newcommand{\lsrmsg}{{\scshape Lsr} message\xspace}
\newcommand{\lsrmsgs}{{\lsrmsg}s\xspace}
\newcommand{\lsumsg}{{\scshape Lsu} message\xspace}
\newcommand{\lsumsgs}{{\lsumsg}s\xspace}
\newcommand{\ackmsg}{{\scshape Ack} message\xspace}
\newcommand{\ackmsgs}{{\ackmsg}s\xspace}

\newcommand{\NN}{
    \ensuremath{%
        \mathop{\rm I\mkern-2.5mu N}%
        \nolimits%
    }%
}

\newcommand{\true}{\keyw{true}}
\newcommand{\false}{\keyw{false}}

\newcommand{\ie}{i.e.\ }
\newcommand{\eg}{e.g.\ }

\newcommand{\awn}{AWN\xspace}
\newcommand{\tawn}{T-AWN\xspace}
\newcommand{\tawnopt}{(T-)AWN\xspace}
\renewcommand{\uppaal}{Uppaal\xspace}

\newcommand{\now}{\keyw{now}\xspace}

\title{Formal Models of the  OSPF Routing Protocol}
\def\titlerunning{Formal Models for OSPF}
\author{Jack Drury
\institute{Data61, CSIRO, Sydney, Australia}
\email{Jack.Drury@data61.csiro.au}
\and
Peter H\"ofner
\institute{Research School of Computer Science\\ANU, Canberra, Australia}
\institute{Data61, CSIRO, Sydney, Australia}
\institute{Computer Science and Engineering\\ UNSW, Sydney, Australia}
\email{Peter.Hoefner@anu.edu.au}
\and
Weiyou Wang
\institute{Data61, CSIRO, Sydney, Australia}
\email{Weiyou.Wang@data61.csiro.au}
}

\def\authorrunning{J. Drury, P. H\"ofner \& W. Wang}
\begin{document}
\maketitle

\begin{abstract}
We present three formal models of the OSPF routing protocol. 
The first two are formalised in the timed process algebra \tawn,
which is not only tailored to routing protocols, but also 
specifies protocols in pseudo-code that is easily readable.
The difference between the two models lies in the level of detail (level of abstraction). 
From the more abstract model we then generate the third model.
It is based on networks of timed automata and can be executed in the model checker \uppaal.

\end{abstract}
\section{Introduction\label{sec:introduction}}
Finding routes between nodes within networks is one of the most common tasks in networking. 
To solve this problem, researchers and engineers have designed dozens of different  routing protocols, which specify how routers communicate with each other and distribute information that enables them to select routes between any two nodes. 
Even though most of these protocols are based on `simple' techniques such as Dijkstra's shortest path algorithm~\cite{Dijkstra59} or the Bellman-Ford algorithm\footnote{%
First proposed by Shimbel~\cite{Shimbel55}; later Bellman~\cite{Bellman58}, Ford~\cite{Ford56} and Moore~\cite{Moore57} published the same algorithm.},
it seems incredibly hard to ensure that the protocols are functionally correct. For example, routing protocols regularly  establish non-optimal routes~\cite{MK10}, the protocol AODV can yield routing loops~\cite{MSWIM13}
or the Border Gateway Protocol (BGP) can exhibit persistent route oscillations~\cite{VaradhanetAl00}.

These examples show that careful protocol analysis is essential. 
However, a formal analysis needs to be based on an unambiguous model, which often does not exist as ``despite the maturity of formal description languages and formal methods for analyzing them, the description of real protocols is still overwhelmingly informal''~\cite{Zave11}.

We present formal and unambiguous models for the Open Shortest Path First (OSPF) routing protocol~\cite{rfc2328,rfc5340}, a widely used  interior gateway protocol. While we are not the first to create formal models for OSPF (see \autoref{sec:related}), it is our belief that we present the first formal model that covers not only the core functionality of OSPF, but also most other details defined in the protocol standard (\autoref{sec:modelI}). 
As usual, that standard~\cite{rfc2328,rfc5340} is written in English prose and contains ambiguities, which we had to resolve. 

Our detailed model is written in the process algebra \tawn~\cite{ESOP16,tawn_tr}, which we briefly describe in \autoref{sec:tawn}. It is a variant of standard process algebras, such as CCS \cite{Mi89}, CSP~\cite{Ho85} or ACP \cite{BK86}.
As for any process algebra, \tawn's semantics is completely formal and hence our model is absolutely precise (no contradictions, no under-/over-specifications) and free of ambiguities.
Moreover, \tawn is particularly tailored for routing protocols such as OSPF and defines the protocol in pseudo-code  that is easily readable by any network or software researcher/engineer.

Modelling many aspects and details of a routing protocol is important, but for an (initial) formal analysis a more abstract model can be extremely useful. \pagebreak[3]
For example, it allows an analysis using model checking techniques, which is infeasible with too many details present, due to state space problems. 
To this end we present, in \autoref{sec:modelII}, a second model, also written in \tawn, that abstracts from aspects such as the retransmission of lost messages, the periodic refreshing of link information and the internal hierarchy of subnets.
Last but not least, in \autoref{sec:modelIII}, we  translate that model into a network of timed automata, which can be executed by the model checker \uppaal~\cite{LPY97,uppaal04}. All three models are presented in full in the appendices.

\section{The Open Shortest Path First (OSPF) Routing Protocol\label{sec:ospf}}

The Open Shortest Path First (OSPF) protocol~\cite{rfc2328} is a widely used proactive, link-state routing protocol used to distribute routing information throughout a single autonomous system.
Being a link-state protocol means that the routers (nodes in the network) exchange topological information about one-hop links with their neighbours, \ie the nodes  within transmission range. That information is flooded through the network so that (eventually) every router has a complete picture of available links in the network. 
This image is used to calculate the shortest/best routes between any two nodes, usually using a variant of Dijkstra's algorithm~\cite{Dijkstra59}. 

OSPF routers use \emph{\hellomsgs} to discover neighbouring nodes. \hellomsgs are broadcast at regular intervals by every node and contain,  next to the sender's identification, a list of all those nodes that the sender has received \hellomsgs from. 
These messages are not flooded through the network, but are sent only a single hop from their origin.

A node will distribute information about the state of its connections by sending \emph{Link State Advertisements} ({\sc Lsa}s). 

Data received by a router that is used for determining the network topology is stored in the \emph{Link-State Database (LSDB)}. 
This database represents the router's current view of the network topology; it contains the most recently received \lsadvs from each unique originator. 
Next to the LSDB, every node maintains a list of discovered neighbours, with whom routing information may be exchanged.

Network nodes also use \hellomsgs (or lack thereof) to determine whether neighbours have become inactive or lost connectivity.
When a node receives the first \hellomsg from a neighbour, it learns of that neighbour's existence. 
If the node finds its own IP address listed within that \hellomsg---meaning that the neighbour is aware of the node's existence---it checks whether it needs to form an adjacency with that neighbour. 

The term \emph{adjacency} describes the relationship between two neighbouring nodes that must exchange all of their topological information---the content of their LSDBs. Eventually, they will have an identical understanding of the network. 
Not all one-hop neighbours will form an adjacency. This is because network traffic can be greatly reduced by forming as few adjacencies as possible whilst still ensuring that all single-hop neighbours will end up with synchronised LSDBs (through a `chaining' of adjacencies). 
When two nodes recognise that they need to become adjacent one of them will be nominated as \emph{master}, the other as \emph{slave}.  It is the master that initiates the exchange of data by sending the necessary information using \emph{Database Description ({\sc Dbd}) messages}. 
The  slave sends \dbdmsgs in response to \dbdmsgs from the master.

Once a node has received a full description of the other node's LSDB, it compares that description with its own LSDB. 
To resolve inconsistencies between the two LSDBs, such as missing entries, it sends \emph{Link State Request} ({\sc Lsr}) messages,
asking for \lsadvs containing the newest information available.
\lsrmsgs are answered by \emph{Link State Update} ({\sc Lsu}) messages; they contain the requested \lsadvs.
Each receipt of an \lsumsg is acknowledged by a \emph{Link State Acknowledgement} ({\sc Lsack}) message.

\emph{Designated routers} are used in combination with adjacency to further reduce network traffic.  
As they are not crucial in the understanding of the main functionality of OSPF we omit a detailed description.

Our model follows closely the OSPF specification as described in the RFC version 2~\cite{rfc2328};
the amendments described in~\cite{rfc5340} are taken into account as well. 
However, many of these changes concern the introduction of IPv6 support---the original specification only allowed IPv4---; as we model IP addresses as arbitrary unique identifiers the amendments had little effect on our model.

\section[]{The Specification Language \tawn\label{sec:tawn}}
One of the standard tools to describe interactions, communications and synchronisations between a collection of agents, processes or network nodes are process algebras.
They provide algebraic laws that allow formal reasoning.
We chose to model OSPF using \tawn~\cite{ESOP16,tawn_tr}, a \emph{timed} process algebra specifically tailored for wireless networks in general and routing protocols in particular. 

\tawn `extends' the (untimed) language \awn (Algebra of Wireless Networks)~\cite{ESOP12},  by timing constructs. In fact, the syntax of both algebras is (nearly) identical.
Both \awn and \tawn provide the right level of abstraction to model key protocol features, while abstracting from implementation-related details.
As its semantics is completely unambiguous, specifying a protocol in such a framework enforces total precision and the removal of any ambiguity. 
{\tawnopt} is tailored for modelling and verifying routing and communication protocols and therefore offers primitives such as {\bf unicast} and {\bf multicast}/{\bf groupcast};%
it defines the protocol in a pseudo-code  that is easily readable---the language itself is implementation independent. 
Currently the tool support is limited; that is why we present an executable model~in~\autoref{sec:modelIII}.

\tawnopt is a variant of standard process algebras (\eg \cite{Ho85,Mi80,BK86}) extended with a local broadcast mechanism and a novel \emph{conditional unicast} operator---allowing error handling in response to failed communications while abstracting from link layer implementations of the communication handling---and incorporating data structures with assignments; its operational semantics is defined in \cite{ESOP12}.

We use an underlying data structure (described later) with several types, variables ranging over these types, operators and predicates. 
First order predicate logic yields terms (or \emph{data expressions}) and formulas to denote data values and statements about them.
The data structure has to contain the types \tDATA, \tMSG, {\tIP} and $\pow(\tIP)$  of \emph{application layer data}, \emph{messages}, \emph{identifiers} and \emph{sets of identifiers}.
The messages comprise \emph{data packets}, containing application layer data, and \emph{control messages}.
A network is modelled as a parallel composition of network routers. Several processes may run in parallel on a single router.

An entire network is modelled as an encapsulated parallel composition of network nodes.

Nodes can only communicate with their direct neighbours, \ie with nodes that are currently within transmission range. 
There are three different ways for internode communication: broadcast, unicast, or an iterative unicast/multi\-cast (called \emph{groupcast} in \tawnopt).

The \emph{process expressions} are given in \autoref{tb:procexpr}.
 \begin{table}[t]
\caption{process expressions~\cite{DIST16}\label{tb:procexpr}}
\vspace{-1ex}
 \centering
{\small
  \setlength{\tabcolsep}{2.6pt}
 \begin{tabular}{|@{\ \ }l@{\ \ }|@{\ \ }p{9.5cm}|}
\hline
\rule[6.5pt]{0pt}{1pt}%
$X(\dexp{exp}_1,\ldots,\dexp{exp}_n)$& process name with arguments\\
$\p+\q$ & choice between process $\p$ and $\q$\\
$\cond{\varphi}\p$&conditional process (if-statement)\\
$\assignment{\keyw{var}:=\dexp{exp}}\p$&assignment followed by process $\p$\\
$\broadcastP{\dexp{ms}}.\p $&broadcast \dexp{ms} followed by $\p$\\
$\groupcastP{\dexp{dests}}{\dexp{ms}}.\p$&iterative unicast or
  multicast to all destinations \dexp{dests}\\
$\unicastP{\dexp{dest}}{\dexp{ms}}.\p \prio \q$& unicast $\dexp{ms}$ to $\dexp{dest}$; if successful proceed with $\p$; otherwise with $\q\hspace{-2.5pt}$\\
$\send{\dexp{ms}}.\p$&synchronously transmit \dexp{ms} to  parallel process on same node\\
$\deliver{\dexp{data}}.\p$&deliver data to application layer\\
$\receive{\msg}.\p$&receive a message\\
\hline
\rule[6.5pt]{0pt}{1pt}%
$\xi,\p$         &process with valuation\\
$V\parl W$	&parallel procs.\ on the same node\\
\hline
\rule[6.5pt]{0pt}{1pt}%
$id\mathop{:}V\mathop{:}R$  & node $id$ running $V$ with range $R$\\
$N\|M$		&parallel composition of nodes\\
$[N]$		&encapsulation\\
\hline
\end{tabular}}
\vspace{-3.5pt}
\end{table}
They should be understandable without further explanation; we add a short description in \autoref{app:AWN}.

When designing or formalising a protocol in \tawnopt, an engineer should not be bothered with timing aspects; except for functions and procedures that schedule tasks depending on the current time. 
Because of this, the only difference between the syntax of \awn and the one of \tawn is that the later is equipped with a local timer \now.

\tawn assumes a discrete model of time, where each sequential process maintains the local variable \now holding its local clock value---an integer. 
Only one clock for each sequential process is employed. All (sequential) processes in a network synchronise in taking time steps, and at each time step all local clocks are incremented by one time unit.  
For the rest, the variable \now behaves as any other variable maintained by a process: its value can be read when evaluating guards, thereby making progress time-dependent, and any value can be assigned to it.
\tawn does not model clock drift nor clock skew implicitly: however, as {\now} is a standard variable that can be changed, these concepts could be integrated in a model. 
As neither of the two concepts is of practical relevance for {\OSPF} we ignore them.

\section{A Detailed Model of OSPF\label{sec:modelI}}
Our first model, written in the specification language \tawn (\autoref{sec:tawn}), is a detailed rendering of OSPF as described in the \emph{Request For Comments} (RFC)~\cite{rfc2328,rfc5340}, the de-facto standard; only a few abstractions are made.
The model is about $165$ lines long, split over $9$ processes, and uses around $30$ functions. 
Compared to the $244$ pages of the RFC, this is a significant reduction in size, while being more precise.
We do not advocate to abandon a specification written in English prose as it often describes the intention and the intuition of protocol designs. However, we believe that many problems of protocol development and specification could be avoided if formal protocol descriptions, such as the one presented here, would accompany the textual specification.

In this section, we describe the abstractions and deviations from the RFC taken, 
present the overall structure of our model and discuss in detail the \tawn process {\HELLO}, which models all the actions after the receipt of a \hellomsg.
The full model is given in \autoref{app:modelI}.

\subsection{Abstractions \& Deviations}

In OSPF only certain pairs of nodes will form adjacencies---not to be confused with node connectivity---and share topological information. 
Although the procedure to find these pairs varies with the capabilities of the network in question, such determination always relies on information provided by a network administrator. As we cannot model the intention of a network administrator, we assume the existence of a predicate $\adj\ip{\ip'}$, 
which evaluates to true if the nodes {\ip} and $\ip'$ are supposed to form an adjacency.

To enhance scalability, OSPF allows networks to be split into multiple \emph{areas}. 
Some of the nodes that are connected to at least two areas will summarise and share intra-area topological information between the adjoining areas. 
Usually, the introduction of areas greatly reduces the number of messages sent.
For the moment we assume a single area; we plan to add this feature in the future.
\pagebreak[3]

\noindent On top of these abstractions we also make the following assumptions.
\begin{enumerate} [label=(\alph*)]
	\item  We assume that every message has a payload of arbitrary size.
	In reality, as the content of LSDBs can be very large, a node splits the information into smaller fragments before sending and the receivers reassemble these fragments. However, this feature is independent of the core functionality of OSPF.
	\item Messages are complete and correct and do not become corrupted (we do allow packet loss, though).
	\item Our model focuses on the sharing of \lsadvs and synchronisation of LSDBs. Since \lsadvs are only shared via bidirectional connections, we assume all connections to be bidirectional.
	\item We abstract from the so-called fight-back mechanism (Sect.\ 13.4 of \cite{rfc2328}), which handles the receipt of self-originated \lsadvs.
\end{enumerate}
\vspace{-2.5pt}

\subsection{Overall Structure\label{sec:structure}}
The detailed model consists of the 9 processes {\OSPF}, {\HELLO}, {\DBD}, \keyw{SNMIS}, \keyw{REQ}, {\UPDp}, \keyw{ACK}, {\QMSG} and {\QSND}.
\begin{itemize}
	\item The main process {\OSPF} reads a single message from the input queue. 
Depending on the type of the message it calls other processes, such as {\HELLO}. 
The process also sends \hellomsgs at periodic intervals and maintains the node's LSDB by removing dead neighbours.
	\item The process {\HELLO} describes all the actions performed when a \hellomsg is received. 
This includes updating the relevant inactivity timer and, if the \hellomsg is from a previously unknown neighbour, updating the node's {neighbour list}. 
\item The process {\DBD} handles incoming \dbdmsgs. 
Sending requests for those \lsadvs from \dbdmsg that do not have a matching partner in the node's LSDB
is among its actions.
	\item The process {\SNMIS} describes the actions required when a sequence number mismatch occurs during the adjacency-establishment procedure. 
\item The process {\REQ} describes the actions following the receipt of an \lsrmsg, such as
 finding the requested \lsadvs in the node's local data, and sending them back to the sender of the message.
\item The process {\UPDp} manages incoming \lsumsgs, including the installation of  up-to-date \lsadvs in the node's LSDB, 
and broadcasting the updated information.	
\item The process {\ACK} describes the actions taken to handle the receipt of an \lsadv. 
\item The process {\QMSG} models a message queue. As before we have two instances, a queue for incoming messages and one for 
outgoing processes.
\item The process {\QSND} is responsible for sending outgoing messages that were generated by the other processes described above.
\end{itemize}
\vspace{-2.5pt}

\subsection{Handling {\scshape Hello} Messages}

The process {\OSPF} handles the receipt of a message using the expression $\receive{\msg}$. 
In case the received message is a \hellomsg, which is tested by the command $\cond{msg=\hello{\ips}{\sip}}$, the process {\HELLO} is called.
Here, $\sip$ is the unique identifier of the sender (usually its IP address) and $\ips$ is the actual content of the \hellomsg, listing all one-hop neighbours of {\sip}.

Next to the content of the received \hellomsg ({\ips} and {\sip}), {\HELLO} maintains the following data:
the node's own identifier {\ip}, 
a set {\nbrs} of \emph{neighbour structures}, 
the Link-State Database {\lsdb} (see \autoref{sec:ospf}) and 
a time value {\hellot} that indicates the time at which {\ip} should send the next \hellomsg---remember that \hellomsgs are sent periodically. 

Each {neighbour structure} stored in {\nbrs} contains, next to the identifier {\nip} of the neighbour, additional information such as an inactivity timer.
The inactivity timer indicates when {\ip} should consider a node {\nip} inactive, and hence when the entry concerning {\nip} should be removed from {\nbrs}.
There are seven other fields that we only discuss in the appendix.
\pagebreak

{\HELLO} first checks whether the sender {\sip} of the \hellomsg is already known,\ie whether the node has received a {\hellomsg} sent by {\sip} before.
This check is performed using the predicate {\fnnbrExist}, which takes the set {\nbrs} of neighbour structures and the sender identifier {\sip} as parameters. 

If {\sip} is unknown (Line~\ref{hello_det_19}) the sender is added to the list of known neighbours {\nbrs}, using the function {\fnnewNBR} (Line~\ref{hello_det_20}). 
Afterwards the process determines what to do, by calling the process {\HELLO}---as the predicate $\nbrExist\nbrs\sip$ now evaluates to true, the actions between Lines~\ref{hello_det_2} and \ref{hello_det_18} are performed.

\setcounter{algorithm}{8}
\vspace{-2pt}
  \algsetup{linenodelimiter=.,linenosize=\tiny}
  \begin{algorithm}[H]
    {
     \caption{Handling \hellomsgs}
      \label{pro:detailed_hello}
      \begin{algorithmic}[1]
\DEFPROCESS{\HELLO}{\ips\comma\sip\comma\ip\comma\nbrs\comma\lsdb\comma\hellot}
	\IF[the sender {\sip} is unknown]{$\neg\nbrExist\nbrs\sip$}																\label{hello_det_19}
		\UPD{\nbrs := \newNBR{\nbrs}{\sip}}																							\label{hello_det_20}
		\helloL\ips\sip\ip\nbrs\lsdb\hellot																								\label{hello_det_21}
	\ELSIF[{\sip} is a known neighbour]{$\nbrExist{\nbrs}{\sip}$}													\label{hello_det_1}
		\UPD{\nbrs := \setINACTT{\nbrs}{\sip}{\now+\rtdeadintvl}}															\label{hello_det_2}
		\UPD{\ns := \getNS{\nbrs}{\sip}}																									\label{hello_det_3}
		\PAR																																			
		\IF[2-WayReceived, start adjacency-forming]{$\ip\in\ips \land \adj\ip\sip \land \ns = \str{Init}$}		\label{hello_det_5}
			\UPD{\nbrs := \setNS{\nbrs}{\sip}{\str{ExStart}}}																		\label{hello_det_6}
			\UPD{\nbrs := \incDDSQN\nbrs\sip}																						\label{hello_det_7}
			\UPD{\nbrs := \setDDT{\nbrs}{\sip}{\now + \rxmtintvl}}															\label{hello_det_8}
			\sendL {\sndmsg{\genDBD{\nbrs}{\lsdb}{\sip}{\ip}}{\{\sip\}}}\ .\ \ospfP{\ip}{\nbrs}{\lsdb}{\hellot}	\label{hello_det_9}
		\ELSIF[Adjacency-forming already underway]{$\ip\in\ips \land \adj\ip\sip \land \ns\geq\str{ExStart}$}							\label{hello_det_10}
			\ospfL\ip\nbrs\lsdb\hellot																										\label{hello_det_11}
		\ELSIF[2-WayReceived, adjacency-forming not needed]{$\ip\in\ips \land \neg\adj\ip\sip$}			\label{hello_det_12}
			\UPD{\nbrs := \setNS{\nbrs}{\sip}{\str{2-Way}}}																		\label{hello_det_13}
			\ospfL{\ip}{\nbrs}{\lsdb}{\hellot}																								\label{hello_det_14}
		\ELSIF[1-WayReceived]{$\ip\notin\ips$}																						\label{hello_det_15}
			\UPD{\nbrs := \initNBR{\nbrs}{\sip}{\str{Init}}}																			\label{hello_det_16}
			\ospfL{\ip}{\nbrs}{\lsdb}{\hellot}																								\label{hello_det_17}
		\ENDIFii	
		\ENDPAR																																	\label{hello_det_18}
	\ENDIFii

		\end{algorithmic}
    }
  \end{algorithm}

\vspace{-4pt}
If the neighbour exists the process first resets the inactivity timer for {\sip} in Line~\ref{hello_det_2}: the value of $\now+\rtdeadintvl$ identifies the latest time when 
the {\ip} should have received another {\hellomsg} from {\sip}.
In Line~\ref{hello_det_3}, the \emph{neighbour state} for node {\sip} is distilled from {\nbrs} and stored in the local variable {\ns}. 
The neighbour state is one of the additional pieces of information in the neighbour structure.
It is a string representing the adjacency status of {\ip} and {\sip}; its value is one of the following.
\begin{description}[parsep=2.5pt,itemsep=0pt]
	\item[Init:] {\ip} has received a \hellomsg from {\sip}, but  a bidirectional communication is not yet verified; 
	\item[2-Way:] communication has been identified as bidirectional, but {\fnadj} does not allow adjacency.  
\end{description}
\noindent The remaining values indicate different stages during the creation of an adjacency between {\sip} and {\ip}.
\begin{description}[parsep=2.5pt,itemsep=0pt]
	\item[ExStart:] the first step of the adjacency-establishment procedure, which decides which router is the master, and which is the slave;
	\item[Exchange:] the two routers {\ip} and {\sip} exchange descriptions of their entire LSDBs, using \dbdmsgs;
	\item [Loading:] nodes compare the received LSDB description with their own and send out \emph{\lsrmsgs} to find missing information, collect newer information or to resolve conflicting information;
	\item [Full:] adjacency is established between {\ip} and {\sip}; the nodes now have identical LSDBs.
	\end{description}
\pagebreak[3]
Depending on the current status between {\ip} and {\sip}, the process {\HELLO} performs different actions.

Let us first consider the case where the receiver's {\ip} is listed in the incoming \hellomsg (\mbox{$\ip\in\ips$}).
Lines \ref{hello_det_6} to \ref{hello_det_9} are executed when the two routers want to form an adjacency, but the process has not yet begun (the status of {\ns} is \str{Init}). In this case, the process 
updates the value of {\ns}  to \str{ExStart} (Line~\ref{hello_det_6}), 
increments \emph{the Database Description Sequence Number} (DDSQN), which is used to ensure that both nodes correctly communicate a description of their entire LSDB (Line~\ref{hello_det_7}), 
resets the timer for the retransmission of \dbdmsgs (Line \ref{hello_det_8}), 
generates and sends a {\dbdmsg} (Line \ref{hello_det_9}) and then returns to the main process {\OSPF}.
For the sake of readability we do not detail how \dbdmsgs are created. 
However, it is worth mentioning that all messages generated by processes such as {\HELLO} are not sent directly to nodes in transmission range, 
but they are first passed to another process running on the same node, using the primitive \textbf{send}.
It is that  process that \textbf{broadcast}s, \textbf{groupcast}s and \textbf{unicast}s the message, respectively.
The reason for that design is two-fold: 
First, it reflects reality as on every router there exists an output queue, which 
collects and sends messages, using a lower-level protocol such as CSMA~\cite{IEEE80211}.
Second, sending messages takes time (in reality and in \tawn); if a message would be sent 
from the process {\HELLO} it could potentially block other incoming message (and their effects).

If the nodes {\sip} and {\ip} have already started the adjacency-establishment procedure---the neighbour state is either \str{ExStart}, \str{Exchange}, \str{Loading} or \str{Full}---then 
no further action is required and the process returns to the main process {\OSPF } (Line \ref{hello_det_11}).
 If {\ip} is listed in {\ips} but the two routers do not want to form an adjacency (Line~\ref{hello_det_12}), the neighbour state {\ns} is set to \str{2-Way} (Line~\ref{hello_det_13}). No other action is performed and the protocol returns to the process~{\OSPF}.

The last possible sequence of actions (Lines~\ref{hello_det_16} and~\ref{hello_det_17}) is executed if 
$\ip$ is not listed in {\ips} and therefore the sender of the \hellomsg is not aware of the receiver's existence.
In this case the process sets {\ns} to \str{Init} for the neighbour structure of {\sip} and returns to {\OSPF}. 
If it was the case that prior to the receipt of this \hellomsg the formula $\ns > \str{2-Way}$ evaluates to true, it means that the two nodes had, at the very least, attempted to form an adjacency. Hence Line~\ref{hello_det_16} also wipes any information related to previous adjacency-forming attempts.

\section{An Abstract Model of OSPF\label{sec:modelII}}
The model described above is too complex for automated analyses such as model checking. 
For example, a network consisting of only two nodes needs to exchange 18 messages before reaching a steady state. One of the main reasons for this complexity is that there is a lot of redundancy built into 
OSPF to protect against message loss. Another contributing factor is that fresh \lsadvs are created and distributed regularly, even when no topological changes have occurred. 
If we assume that all sent messages are received\footnote{This is one of the fundamental features of \tawnopt.} we can simplify the model without changing its core functionality.
For example, we no longer need retransmission lists and timers, and we only need to generate \lsadvs in case of a topological change.

The simplified model comes in at roughly $70$ lines, splits over $7$ processes and uses only $12$ functions. 
As before, we only present the handling of \hellomsgs;  the full model is given in \autoref{app:modelII}.%

\subsection{Abstractions \& Deviations}
As this model is an abstraction of the previous model, we make the same assumptions as in \autoref{sec:modelI}.

Additionally, we now assume that all sent messages are received. 
This means that we no longer need to maintain retransmission lists, 
as their sole purpose is to hold messages that are yet to be acknowledged. 
We are also able to remove acknowledgement messages for the same reason.
We assume that all {\lsrmsgs} will be handled and responded to correctly. 
Along with that, we further assume that an {\lsrmsg} can contain multiple \lsadv headers. By doing so, a router can request all \lsadvs needed by sending a single request.
According to the RFC all routers should refresh their \lsadvs at regular intervals by sending out \lsumsgs, even if nothing has changed. We assume that this is no longer necessary and remove this functionality.
Finally, we assume that each node will form adjacencies with all its one-hop neighbours. In terms of exchanging topological information, this is equivalent to assuming that all connections are point-to-point. 

\subsection{Handling {\scshape Hello} Messages}
The overall structure is the same as for the detailed model (\autoref{sec:structure}), with the exception that the processes {\SNMIS} and {\ACK} are not needed any longer.
The simplified process {\HELLO} is shown in \autoref{pro:simple_hello}. 
As we no longer have to check whether nodes form adjacencies the process simplifies drastically.

\setcounter{algorithm}{1}
  \algsetup{linenodelimiter=.,linenosize=\tiny}
  \begin{algorithm}[H]
    {
     \caption{Handling \hellomsgs (simplified)}
      \label{pro:simple_hello}
      \begin{algorithmic}[1]
\DEFPROCESS{\HELLO}{\ips\comma\sip\comma\ip\comma\nbrs\comma\lsdb\comma\hellot}
	\IF[the sender {\sip} is unknown]{$\neg\nbrExist\nbrs\sip$}																				\label{hello_simp_1}
		\UPD{\nbrs := \newNBR\nbrs\sip}																												\label{hello_simp_2}
		\UPD{\nbrs := \setINACTT\nbrs\sip{\now + \rtdeadintvl}}																				\label{hello_simp_3}
		\UPD{\lsa := \newLSA\ip\now\nbrs}\COM{generate new \lsadv}																	\label{hello_simp_4}
		\UPD{\lsdb := \install\lsdb{\{\lsa\}}}\COM{ install \lsadv}																				\label{hello_simp_5}
		\sendL{\sndmsg{\upd{\{\lsa\}}{\ip}}{\{nip \,\lvert\, (nip, *)\in\nbrs\}}}\ . \COM{send \lsadv}								\label{hello_simp_6}
		\sendL{\sndmsg{\dbd{\{\hdr{lsa} \mid lsa\in\lsdb\}}{\ip}}{\{\sip\}}}\ .   \COM{send {\sc Dbd} back to \sip}		\label{hello_simp_7}
		\ospfL\ip\nbrs\lsdb\hellot 																															\label{hello_simp_8}
	\ELSIF[{\sip} is a known neighbour]{$\nbrExist{\nbrs}{\sip}$}																			\label{hello_simp_9}
		\UPD{\nbrs := \setINACTT{\nbrs}{\sip}{\now + \rtdeadintvl}}																		\label{hello_simp_10}
		\ospfL\ip\nbrs\lsdb\hellot 																															\label{hello_simp_11}
	\ENDIFii
		\end{algorithmic}
    }
  \end{algorithm}

The only case distinction that remains is to check whether router {\ip} has seen messages from the sender {\sip} of the \hellomsg (Lines \ref{hello_simp_1} and \ref{hello_simp_9}).
In the case that {\ip} has previously received a message from the sender ($\nbrExist\nbrs\sip$), the inactivity timer is reset (Line \ref{hello_simp_10}) and  the process returns to the {\OSPF} process (Line \ref{hello_simp_11}), ready to receive another message. 
These two lines are the only actions that remain of Lines \ref{hello_det_2}--\ref{hello_det_18} in the detailed \autoref{pro:detailed_hello}.
Otherwise, if the message received is the first \hellomsg from {\sip} ($\neg\nbrExist\nbrs\sip$), the router {\ip} creates a new neighbour structure  (Line \ref{hello_simp_2}) and initialises its inactivity timer (Line \ref{hello_simp_3}).
Taking into account that in the detailed model the process {\HELLO} is called in Line \ref{hello_det_21},  these two lines correspond to Line \ref{hello_det_20} and Line \ref{hello_det_2} of \autoref{pro:detailed_hello}, respectively.
Afterwards, {\ip} generates a new \lsadv (Line \ref{hello_simp_4}), adds it to its own database {\lsdb}  (Line \ref{hello_simp_5}) and sends it out to all of the neighbours it is aware of, which are determined by the formula $\{\nip \,\lvert\, (\nip, *)\in\nbrs\}$; here, all one-hop neighbours will form an adjacency and there is no lengthy establishment procedure, therefore the nodes are considered adjacent upon discovery.

Then, in Line  \ref{hello_simp_7}, the receiving router {\ip} sends a \dbdmsg to the originator {\sip} of the \hellomsg so that they can compare their LSDBs. 
This is similar to Line \ref{hello_det_9} of \autoref{pro:detailed_hello}. 
However, the data structure of the detailed model is more complex and therefore more assignments and different functions are needed.
Finally the process returns to {\OSPF}.
\section{An Executable Model of OSPF\label{sec:modelIII}}
We translate the simplified \tawn model into an executable model, which can be used by \uppaal~\cite{uppaal04,LPY97}, which is an established model checker for \emph{networks of timed automata}.
\uppaal is well suited for an automated analysis of \tawnopt specifications as it provides
(a) two synchronisation mechanisms, which translate to uni- and broadcast communication;
(b) common data structures, such as arrays and structs, and a C-like programming language to define updates on these data structures;
(c) mechanisms for time.

\subsection{\uppaal's Specification Language}
\uppaal accepts networks of timed automata with guards and data structures as input. 
The state of the system  is determined, in part, by the values of data variables that can be either shared between automata, or local. 
As for \tawnopt, we assume a data structure with several types, variables ranging over these types, operators and  predicates. 

Each automaton is a graph, with locations, and edges between locations. 
Every edge has a guard---if not specified the guard is {\tt true}---, optionally a synchronisation label, and an update. 
Synchronisation occurs via so-called channels. For each channel $a$ there is one label $a!$ to denote the sender, and $a?$ to denote the receiver. 
Transitions without labels are internal; all others use one of two types of synchronisation.

In \emph{binary handshake} synchronisation, one automaton having an edge with a label that has the suffix~$!$ synchronises with a single other automaton that has an edge with the same label including a $?$-suffix. 
These two transitions synchronise when both guards are true in the current state, and only then. 
When the transition is taken both locations change, and the updates will be applied to the state variables; first the updates on the $!$-edge, then the updates on the $?$-edge. 
If there is more than one possible pair, then the transition is selected non-deterministically.

In \emph{broadcast} synchronisation, one automaton with a $!$-labelled edge synchronises with a set of other automata. 
The initiating automaton can change its location, and apply its update, if the guard on its edge evaluates to true. 
It does not require a second synchronising automaton. Automata with a matching $?$-labelled edge have to synchronise if their guard is currently true. 
They change their location and update the state. 
The automaton with the $!$-edge will update the state first, followed by the other automata.

\subsection{\tawn to \uppaal}
The translation from \tawnopt to \uppaal is more or less straightforward. 
A possible translation from {\awn} to {\uppaal} is sketched in~\cite{Moehring17}.\footnote{A similar translation is also mentioned in \cite{TACAS12}.} 

The main process {\OSPF} (including the subprocesses it calls, such as {\HELLO}) of every node in the network is modelled as a single automaton, each having its own data structures such as an  LSDB and message buffers. 
A node's output queue is modelled as a separate automaton. 
As a slight optimisation we are able to avoid another automaton for the input queue. A network of $n$ nodes is modelled by $n$ copies of the two automata; each tagged with a unique identifier.

The implementation of the data structure defined in \mbox{\tawn} is straightforward. 
An LSDB for example is an array of LSAs, one entry for every node in the network. An LSA  is given by the data type

\vspace*{-2pt}
\begin{uppaalcode}[basicstyle=\footnotesize,backgroundcolor=\color{backgroundColour},numbers=none]
     typedef struct                    
     { int [0,N-1] ip;          // identifier of originator
       int[0,age_bound+1] age;  // time when the LSA is originated
     } LSA;
\end{uppaalcode}

\vspace*{-4pt}
\noindent where \verb+age+ denotes the time when the LSA was generated.
\pagebreak[3]

\begin{figure}[t]
    \centering
    \begin{subfigure}[t]{0.3\textwidth}
    		\centering
        \includegraphics[scale=.12]{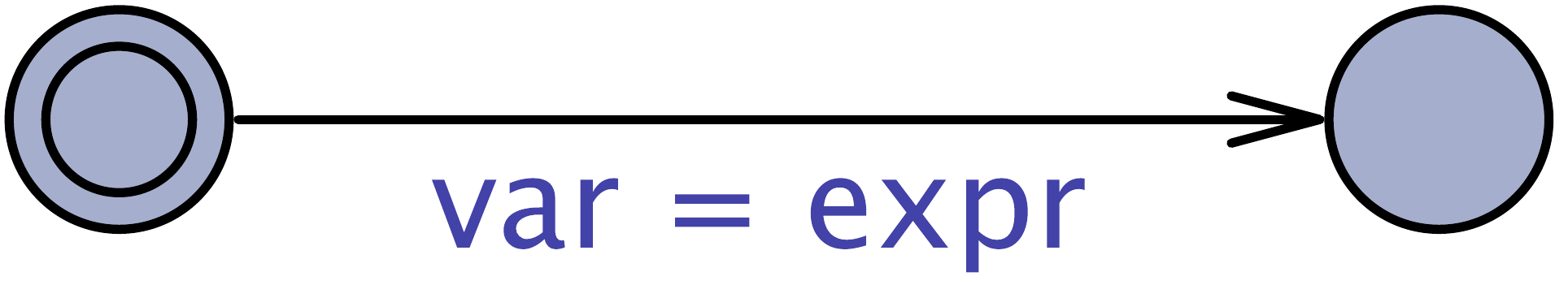}
        \caption{\mbox{$\assignment{\var{var} :=\dexp{expr}}$}}
        \label{subfig_a}
    \end{subfigure}
    \qquad\begin{subfigure}[t]{0.4\textwidth}
    		\centering
        \includegraphics[scale=.17]{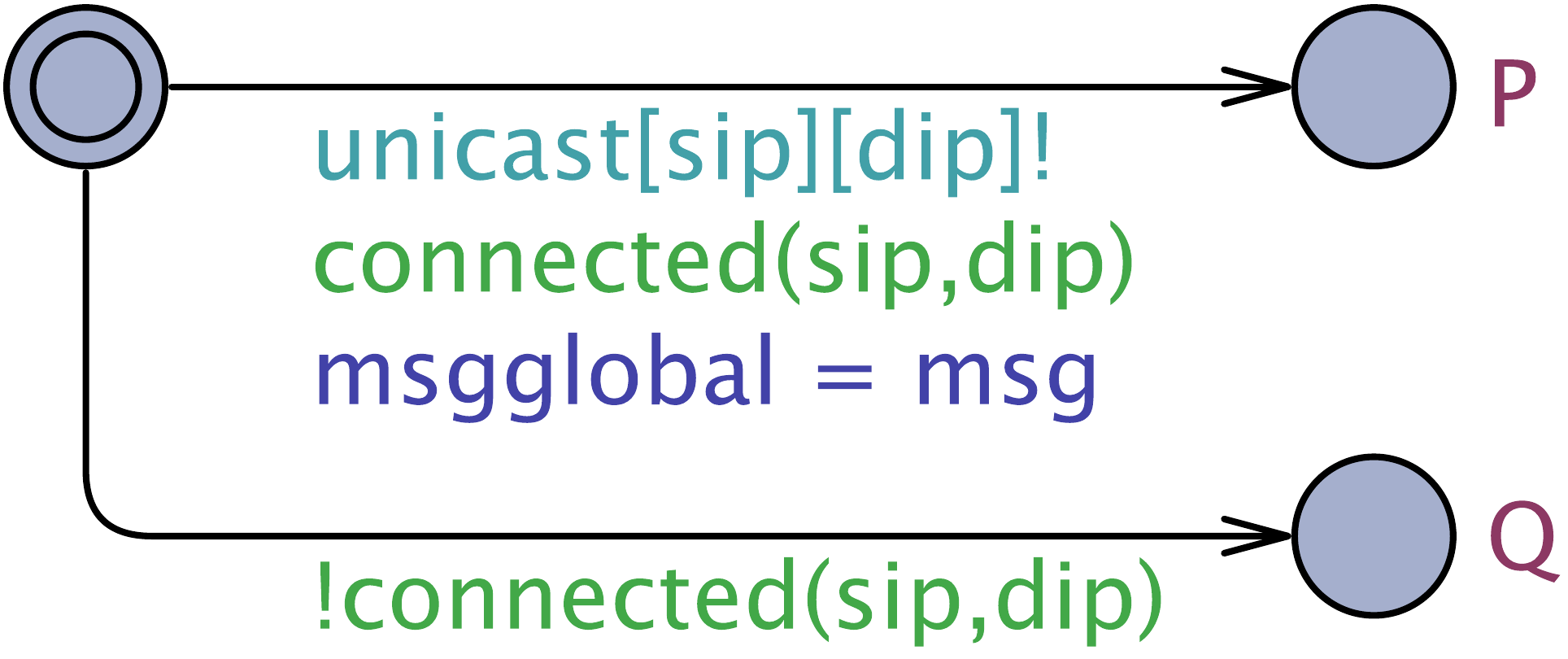}
        \caption{\mbox{$\unicastP{\exp{msg}}{\dexp{dip}}.P\prio Q$}}
        \label{subfig_b}
    \end{subfigure}
\vspace{-4pt}
    \caption{Translating \tawn to \uppaal\label{fig:trans}}
\vspace{-15pt}
\end{figure}

An assignment $\assignment{\var{var} :=\dexp{expr}}$ is translated into an edge with a single update, as depicted in \autoref{subfig_a}. 
Similarly, the translation of a conditional $[\varphi]$ yields an edge with a single guard. 
Translating the sending primitives is more complicated as one has to take care of the intended destinations, the connectivity between nodes as well as the exchange of the actual messages.
 \autoref{subfig_b} depicts the translation schema for \textbf{unicast}: the locations named \verb+P+ and \verb+Q+ stand for sub-automata resulting from the translation of the subprocess $P$ and $Q$; \verb+unicast[sip][dip]+ represents a unique channel where data is sent from node \verb+sip+ to \verb+dip+; the predicate \verb+connected+ characterises the topology---only when the two nodes are connected is the  message transferred. 
When the transition is taken, the sender \verb+sip+ copies its message to a global variable \verb+msgglobal+, and the receiver copies it subsequently to its local buffer.
As we currently do not have software supporting this translation we perform the translation manually; this allows us to optimise the \uppaal model. For example, a sequence of assignments can be combined into a single edge.

A major difference between the \tawnopt- and \uppaal models is that in \uppaal message queues can only be of finite length, which is in contrast to the \tawn model. The local message buffer is modelled as a fixed-length array of messages. \uppaal will give a warning if during model checking an out-of-bounds error occurs, \ie if the array was too small. 
Another difference is that \tawnopt is based on discrete time, while \uppaal uses dense time. It is easy to see that (for our models) an embedding of discrete time into dense time does not cause any problems.

\begin{figure}[h!]
\centering
\includegraphics[width=.87\textwidth]{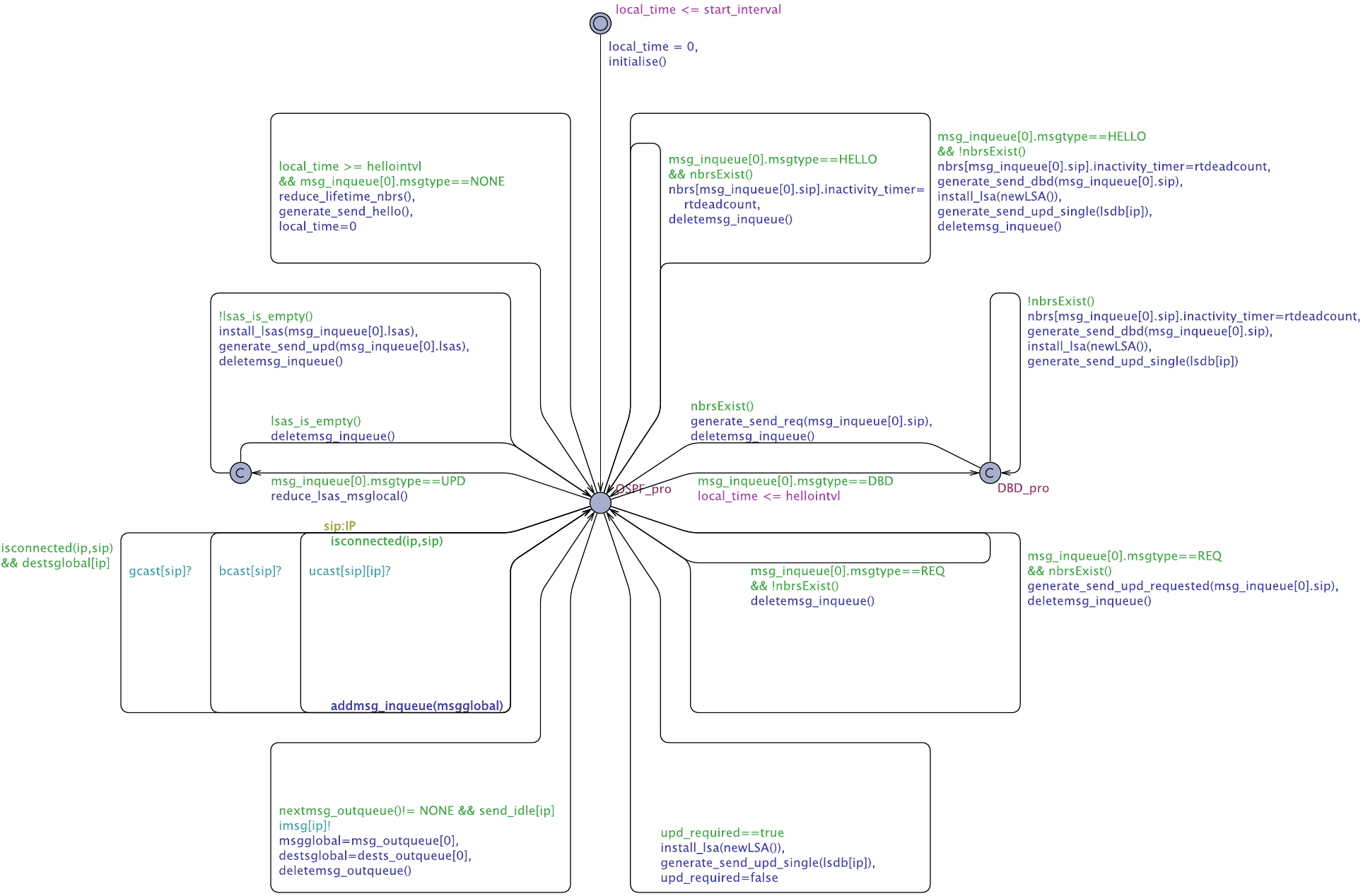}
\vspace{-5pt}
\caption{Executable Model of OSPF as timed automaton\label{fig:uppaal2}}
\vspace{-5pt}
\end{figure}

We translate the simpler model, presented in \autoref{sec:modelII}, into an executable \uppaal model. The resulting timed automaton modelling \OSPF, including its subprocesses such as {\HELLO}, is presented in \autoref{fig:uppaal2}.
\vspace{-16pt}

\newpage
\noindent
We have assumed that each node will form adjacencies with all its one-hop neighbours; an assumption that significantly decreased the size of our model. 
For model checking, however, this assumption potentially increases the state space as creating and maintaining adjacencies requires a lot of message passing.

\subsection{Handling Hello Messages}
\begin{figure}[t]
\centering
\includegraphics[width=.85\textwidth]{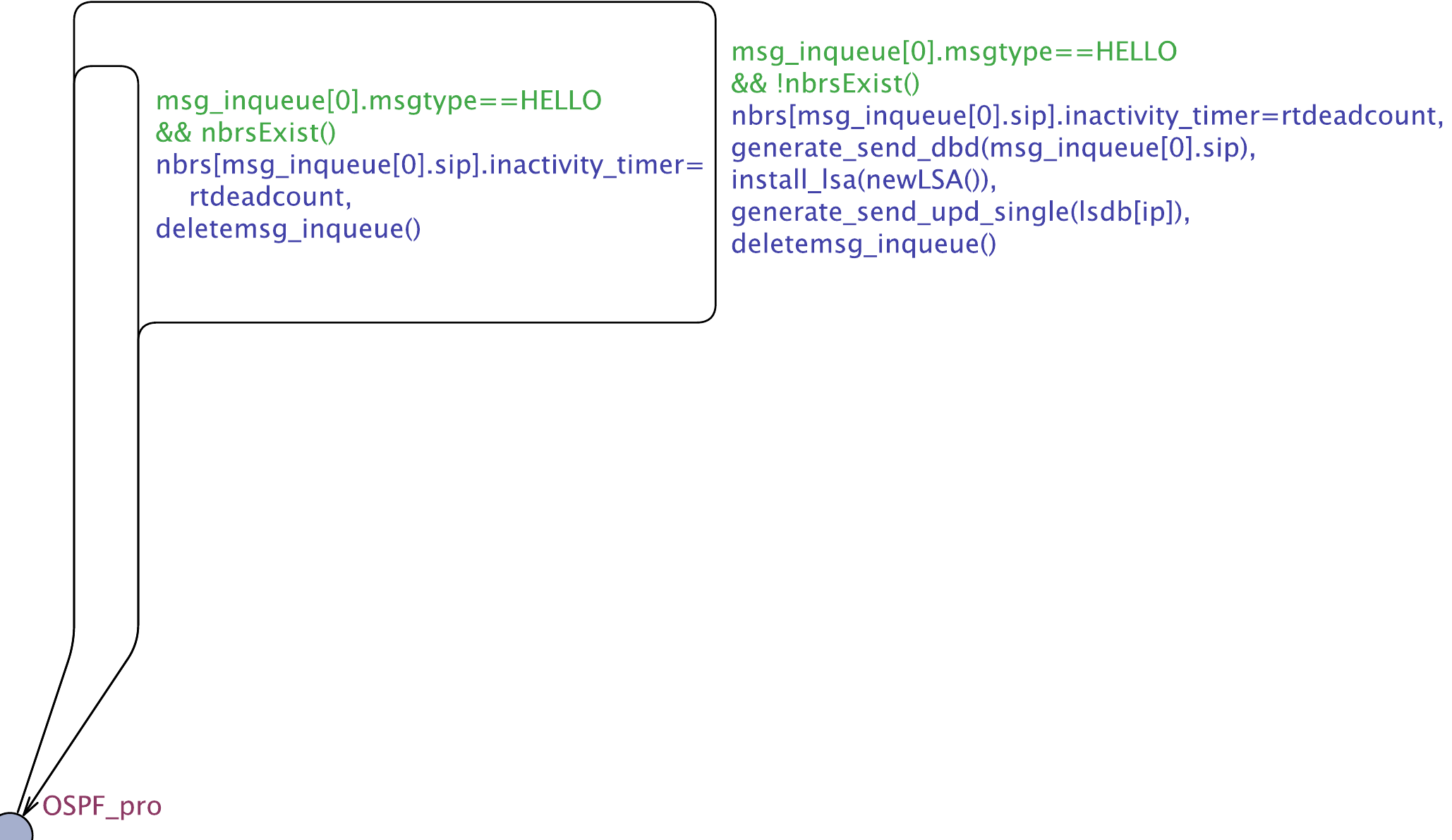}
\vspace{-9mm}
\caption{Translation of the process {\HELLO}\label{fig:uppaal3}}
\end{figure}

To illustrate the relationship between \tawn and \uppaal we present the translation of the {\tt HELLO} process in \autoref{fig:uppaal3}. 
That figure is an enlarged version of the right-upper corner of the automaton in \autoref{fig:uppaal2}. 
For simplicity we skip the data structures and the definitions of functions occurring in the timed automaton; for the full description see \autoref{app:modelIII}.

The two edges correspond with the two conditions in Lines~\ref{hello_simp_1} and \ref{hello_simp_9} of the process {\HELLO} (\autoref{pro:simple_hello}).

The outermost edge---the one that states \verb+&& !nbrsExist()+ in the second line---combines 
Lines~\ref{hello_simp_1}--\ref{hello_simp_8} of \autoref{pro:simple_hello}
in a single edge.
The guard first checks that the received message is indeed a \hellomsg (\verb+msg_inqueue[0].msgtype == HELLO+) and then validates whether it received a \hellomsg from the sender using \verb+&& !nbrsExist()+. The latter is a direct translation of Line~\ref{hello_simp_1}.
The remaining actions of \autoref{pro:simple_hello} are translated into simple updates of the data structure. Additionally we have to delete the received \hellomsg  from the input queue. This is done by \verb+deletemsg_inqueue+. 

The innermost path corresponds directly to Lines~\ref{hello_simp_9}--\ref{hello_simp_11}. 
It describes the scenario where the router has seen a \hellomsg from the sender before.
As before, we need the additional action of deleting the \hellomsg.
\pagebreak[3]

\section{Related Work\label{sec:related}}

For our modelling efforts we use the process algebra \tawn, as well as networks of timed automata that form the input language of the model checker \uppaal.

Next to \tawn, several process algebras modelling broadcast communication have been
proposed, such as
the Calculus of Broadcasting Systems (CBS) \cite{CBS91,CBS},
the $b\pi$-calculus \cite{bpi},
CBS\# \cite{NH06},
the Calculus of Wireless Systems (CWS) \cite{CWS},
the Calculus of Mobile Ad Hoc Networks (CMAN) \cite{CMAN},
the Calculus for Mobile Ad Hoc Networks (CMN) \cite{CMN},
the $\omega$-calculus \cite{SRS10},
restricted broadcast process theory (RBPT) \cite{RBPT},
$bA\pi$ \cite{bApi} and 
the broadcast psi-calculi \cite{BHJRVPP11}.
However, \tawn is the {only} process algebra that provides all  features needed to fully model routing protocols such as OSPF, namely data handling, (conditional) unicast and (local) broadcast. 
Moreover, all other process algebras lack the feature of guaranteed receipt of messages. 
Due to this, it is impossible to analyse properties such as route discovery in process algebras different to \tawn: only in \tawn a failure of route discovery can be interpreted as an imperfection in the protocol, rather than as a result of a chosen formalism not ensuring guaranteed receipt.

There are only a few other formal models for OSPF.
Nakibly et al.~\cite{NakiblyEtAl14} created a model of {\OSPF} for the model checker CBMC~\cite{CBMC}\footnote{CBMC is a bounded model checker accepting a simplified C-language as input.}. 
Their model uses a fixed topology---similar to our model-checking models, but different to our \tawn models---and abstracts from many crucial details such as \hellomsgs and \dbdmsgs.
The online model---a link is given in~\cite{NakiblyEtAl14}---analyses a tiny topology with three nodes.
Another model, that can be fed into the Z3-Solver~\cite{Z3}, is described in~\cite{Malik}. The authors claim that it is a detailed model covering concepts such as designated routers and claim that network topologies up to 30 nodes can be analysed. This would be very impressive, but the model is neither described in detail in~\cite{Malik} nor available online; hence a comparison to our model is impossible.
It is our belief, that the formal models available
are far too abstract to give guarantees about the correctness of OSPF. By providing more models we bridged this gap.
\section{Conclusion and Future Work}
We have presented three different models of the routing protocol OSPF.  
Two are written in the specification language \tawn, a process algebra particularly tailored to routing protocols. 
The first model covers many details of OSPF and, to the best of our knowledge,  is by far the most detailed model available. 
The second model is based on additional assumptions, making it significantly shorter and easier. 
From the simpler model we have derived a model that can be executed in the model checker \uppaal. 

\noindent Future work is three-fold.
\begin{enumerate} [label=(\alph*)]
\item We plan to add additional concepts to our detailed model.
This includes protocol-specific details such as the fight-back mechanism (Sect. 13.4 of \cite{rfc2328}) and a more dynamic definition of adjacency (\fnadj), as well as more general concepts, such as 
 \emph{areas}, which are mentioned in \autoref{sec:modelI} and 
		\emph{interfaces}. The latter may require a small adaptation of \tawn.
\item Moreover, we want to validate our models against real implementations. One of the main problems when 
	creating formal specifications for, or implementations of routing protocols is that they always represent the personal view of the developer(s). 
	This is due to the fact that standards are written in English prose. We plan to automatically compare our \uppaal model with off-the-shelf 
	implementations of OSPF, such as the well-known Quagga implementation~\cite{rfc2328,rfc5340}, running on the network emulator CORE~\cite{Ahrenholz08}.
\item Last, but certainly not least, we are going to analyse OSPF. This analysis will include classic properties such as the avoidance of routing loops, but also security aspects. 
For the latter we have to develop formal attack models.
\end{enumerate}

\vspace{-2mm}
\paragraph{Acknowledgments}
This work was conducted in partnership with the Defence Science \& Technology Group and Data61, CSIRO, through the \emph{Next Generation Technologies Fund}.

\bibliographystyle{eptcs}
\bibliography{ospf}
\appendix
\newpage
\section[]{Informal Description of the Process Expressions of {\tawnopt}\label{app:AWN}\footnotemark}
\footnotetext{This appendix  (in nearly identical form) also appeared in~\cite{MARS17}.}
In this appendix we describe the \emph{process expressions} of \tawnopt, given in \autoref{tb:procexpr}.
 
A process name $X$ comes with a \emph{defining equation}\vspace{-1ex}
\[
	X(\keyw{var}_1,\ldots,\keyw{var}_n) \stackrel{{\it def}}{=} P\,,
\]
where $P$ is a process expression, and the $\keyw{var}_i$ are data variables maintained by process $X$. A named process is like a \emph{procedure}; 
when it is called, data expressions $\dexp{exp}_i$ of the appropriate type are filled in for the variables $\keyw{var}_i$.
Furthermore, $\varphi$ is a condition, $\keyw{var}\mathop{:=}\dexp{exp}$ an assignment of a data expression \dexp{exp} to a variable \keyw{var} of the same type, \dexp{dest}, \dexp{dests}, \dexp{data} and \dexp{ms} data expressions of types {\tIP}, $\pow(\tIP)$, {\tDATA} and {\tMSG}, respectively, and $\msg$ a data variable of type \tMSG.

Given a valuation of the data variables by concrete data values, the process $\cond{\varphi}\p$ acts as $\p$ if $\varphi$ evaluates to {\tt true}, and deadlocks if $\varphi$ evaluates to {\tt false}.%
\footnote{
	As \label{fn:undefvalues} operators we also allow \emph{partial} functions with the convention that any atomic formula containing an undefined subterm evaluates to {\tt false}.}
In case $\varphi$ contains free variables that are not yet interpreted as data values, values are assigned to these variables in any way that satisfies $\varphi$, if possible.
The  process $\assignment{\keyw{var}\mathop{:=}\dexp{exp}}\p$ acts as $\p$, but under an updated valuation of the data variables.
The  process $\p+\q$ may act either as $\p$ or as $\q$, depending on which of the two is able to act at all.  
In a context where both are able to act, it is not specified how the choice is made. 
The process $\broadcastP{\dexp{ms}}.\p$ broadcasts (the data value bound to the expression) $\dexp{ms}$ to all connected components (routers), and subsequently acts as $\p$, whereas the process $\unicastP{\dexp{dest}}{\dexp{ms}}.\p \prio \q$ tries to unicast the message $\dexp{ms}$ to the destination \dexp{dest}; 
if successful it continues to act as $\p$ and otherwise as $\q$.
The process $\groupcastP{\dexp{dests}}{\dexp{ms}}.\p$ tries to transmit \dexp{ms} to all destinations $\dexp{dests}$, and proceeds as $\p$ regardless of whether any of the transmissions is successful.
The process $\receive{\msg}.\p$ receives any message $m$ (a data value of type \tMSG) either from another network node, from another  process running on the same node or from an application layer process connected to that component. 
It then proceeds as $\p$, but with the data variable $\msg$ bound to the value $m$.  
In particular, $\receive{\newpkt{\dexp{id}}{\dexp{data}}}$ models the injection of data from the application layer, where the function $\newpktID$ generates a message containing the application layer $\dexp{data}$ and the identifier $\dexp{id}$, here indicating the message type.
Data is delivered to the application layer by \deliver{\dexp{data}}.

The internal state of a sequential process described by an expression $P$ in this language is determined by $P$, together with a \emph{valuation} $\xi$ associating data values $\xi(\keyw{var})$ to the data variables \keyw{var} maintained by this process.
In case a process maintains no data values, we use the empty valuation $\xi_0$.
A \emph{valuated process} is a pair $(\xi,P)$ of a sequential process $P$ and an initial valuation $\xi$.

Finally, $V\parl W$ denotes a parallel composition of valuated processes $V$ and $W$, with information piped from right to left; in typical applications (\eg ours or~\cite{DIST16}) $W$ is a message queue.

In the full process algebra \cite{ESOP12}, \emph{node expressions} $\dval{id}\mathop{:}V\mathop{:}R$ are given by process expressions $V$, annotated with a (unique) process identifier~$\dval{id}$ and a set of nodes $R$ that are connected to~$\dval{id}$.

A partial network is then modelled as a parallel composition of node expressions, using the operator~$\|$, and a complete network is obtained by placing this composition in the scope of an encapsulation operator $[\_\!\_\,]$. 
The main purpose of the encapsulation operator is to prevent the receipt of messages that have never been sent by other nodes in the network---with the exception of messages $\newpkt{\dexp{data}}{\dexp{dip}}$ stemming from the application layer of a node.

\newpage
\setcounter{algorithm}{0}
\section{The Simple T-AWN Model\label{app:modelII}}
We present all details concerning our simple \tawn model, which we sketched in \autoref{sec:modelII}.
Although we created the detailed model first, we believe that presenting the simple model in detail first is easier for the reader as she can concentrate on the core aspects of OSPF. 
When presenting the detailed model in \autoref{app:modelI} we can concentrate on the differences, which makes the description  shorter.

\newcommand{\spac}{\rule[-9pt]{0pt}{10pt}}
\subsection{Data Structure}
We describe the data structures required for our formal specification of OSPF. 
First we define a number of types that are used to handle and process information.
Based on these types,  we then define functions.
Finally, we describe the exact intention and effects of the various operations performed by each node.

Our formal specification of OSPF focuses on adjacency establishment between two routers and the LSA flooding mechanism. 
As a node receives messages from its neighbours it will update its \emph{neighbour list}, which documents its conversations with other nodes. After new neighbours or other topological changes are discovered, information of the local topology will be updated and  shared in the form of \lsadvs. 
Both the \emph{neighbour structures} and \emph{link state databases} are key components of the data structure. 

\subsubsection{Mandatory Types of \tawnopt}
\textit{Messages} are used to send information via the network. In our specification we use the variable {\msg} of type \tMSG. 
We distinguish four types of OSPF control messages: \hellomsgs, \dbdmsgs, \lsrmsgs, and \lsumsgs.
We also use an internal message for intranode communication.

The type {\tIP} describes a set of IP addresses or, more generally, \textit{a set of node identifiers}. 
We assume that each node has a unique identifier \var{ip} $\in$ \tIP. 
Each node maintains a variable \var{ip}, which always stores the node's identifier. 
Every message contains the variable \var{sip} that holds the sender's identifier. 
Furthermore, \var{nip} denotes an identifier  of an arbitrary neighbour.

Lastly, \tawn  provides the variable {\now} of the type {\tTIME}, which is local to a node and increases whenever time passes.

\subsubsection{Link State Advertisements (\lsadvs) \& Link State Databases}

Link State Advertisements (\lsadvs) are one of the key components of OSPF. They are used to describe how routers and networks are interconnected and will be advertised through parts of the network. 
The RFC defines five distinct types of \lsadvs. 
For the simple model we assume an area with Point-to-Point networks only; thus only Router-\lsadvs exist in our model. 
A \emph{Router-\lsadv} describes the outgoing `links' belonging to a given router, which effectively describes the connections that this router provides. 

Every router periodically generates and distributes a Router-\lsadv (for brevity we refer to it as an \lsadv) to describe its knowledge about the local network topology. 
An \lsadv is given by three components:

\begin{enumerate}
	\item The IP address of the originator, which is an element of \tIP.
	\item A timestamp denoting when the {\lsadv} was generated,  which is an element of \tTIME.\footnote{%
			We assume the type {\tTIME} to be isomorphic to the integers, which allows us to use operations such as addition or subtraction.}
	\item A set of IP addresses representing all outgoing links of the originator (discovered so far), which is an element of $\pow(\tIP)$.
\end{enumerate}
\noindent We denote the type of \lsadv by {\tLSA} and define a function $\keyw{(\_,\_,\_)} : \tIP \times {\tTIME} \times \pow(\tIP) \rightarrow {\tLSA}$
to generate \lsadvs.
An \lsadv is uniquely identified by its first two components, the \emph{\lsadv header}. 
The type of headers is denoted by ${\tLSAHDR}$.
We define a generator function
$
	\keyw{(\_,\_)} : \tIP \times {\tTIME} \rightarrow {\tLSAHDR}
$
and an extraction function to get the header from an \lsadv: \\[2mm]
\centerline{$\begin{array}{rcl}
	\fnhdr : {\tLSA} &\rightarrow& {\tLSAHDR}\\
	\hdr{\lsa}&=& (\pi_1(\lsa), \pi_2(\lsa))\,.\footnotemark
\end{array}$}\\[2mm]
\footnotetext{We use projection functions $\pi_{i}$ to distill the $i$-th component of a tuple.}%
When the protocol encounters two \lsadvs with the same originator, it needs to decide which one is more recent. 
Therefore, we define a partial order on {\tLSAHDR}.\\[2mm]
\centerline{$\begin{array}{rcl}
	\lsahdr_1 \leq \lsahdr_2 &:\iff& \pi_1(\lsahdr_1) = \pi_1(\lsahdr_2) \land \pi_2(\lsahdr_1) \leq \pi_2(\lsahdr_2)
\end{array}$}\\[2mm]
As usual, we define a strict partial order by $\lsahdr_1<\lsahdr_2 :\iff \lsahdr_1\leq \lsahdr_2\land \lsahdr_1\not=\lsahdr_2$.
The term $\hdr{lsa_1} < \hdr{lsa_2}$ indicates that $lsa_2$ is more recent than $lsa_1$ ($lsa_2$ was generated later) and should replace $lsa_1$ in the link state database to keep it up to date.

A \emph{link state database} is a set of \lsadvs, where each \lsadv is uniquely identified by its originator. The type, denoted by {\tLSDB}, is defined as\\[2mm]
\centerline{$
	{\tLSDB} := \{ lsdb \mid lsdb \in \pow({\tLSA}) \>\> \land \>\> \forall\, lsa_1, lsa_2 \in lsdb : lsa_1 \neq lsa_2 \Rightarrow 
	\hdr{lsa_1} \not\leq \hdr{lsa_2} \}\,.$}\\[2mm]
Every router maintains a link state database and uses it to calculate the routing table. 
 models do not consider routing table calculations and only focus on synchronisation of link state databases.
Note that this data type is equivalent to\\[2mm]
\centerline{$\{ lsdb \mid lsdb \in \pow({\tLSA}) \>\> \land \>\> \forall\, lsa_1, lsa_2 \in lsdb : lsa_1 \neq lsa_2 \Rightarrow 
	\pi_1(lsa_1) \neq \pi_1(lsa_2)\}\,.$}

\vspace{-2mm}
\subsubsection{Data Structure Relating Neighbours}
A router stores a set of {neighbour structures} so that it can keep track of how much and whether it needs to exchange information with each neighbour. 
Since we assume point-to-point networks it means that routers must exchange information with all of their neighbours.
This fact simplifies the structure greatly when compared to the detailed model.

\noindent A \emph{neighbour structure} contains two components:
\begin{enumerate}
	\item The IP address of the neighbour, which is an element of \tIP.
	\item A timestamp of type {\tTIME}, which indicates when the neighbour is considered inactive or dead---unless another message is received from this neighbour before that time.
\end{enumerate}
{\tNBR} denotes the type of neighbour structures and an element is generated by $\keyw{(\_,\_)} : \tIP \times {\tTIME} \rightarrow {\tNBR}$.

A \emph{neighbour list} is a set of neighbour structures, where each IP address has at most one entry. 
These correspond to the Interface Data Structure in the RFC.\\[2mm]
\centerline{$
	{\tNBRS} := \{ nbrs \mid nbrs \in \pow({\tNBR}) \>\> \land \>\> \forall\, nbr_1, nbr_2 \in nbrs : nbr_1 \neq nbr_2 \Rightarrow \pi_1(nbr_1) \neq \pi_1(nbr_2) \}\,.$}

\vspace{-2mm}
\subsubsection{Messages}
The simple model uses four types of OSPF messages:
\begin{enumerate}
	\item A \hellomsg contains two entries: (a) a set of IP addresses of all neighbours discovered by the sender and (b) the identifier of the sender itself.
	The generation function is
	
\vspace{1.2mm}
\centerline{$\helloID : \pow(\tIP) \times \tIP \rightarrow {\tMSG}\,.\spac$}

	\item A \dbdmsg consists of two fields: (a) a set of \lsadv headers describing the LSDB of the sender (an element of $\pow({\tLSAHDR})$) and (b) the sender itself.	
	The generation function is
	
\vspace{1.2mm}
\centerline{$\dbdID : \pow({\tLSAHDR}) \times \tIP \rightarrow {\tMSG}\,.$
}

	\item An \lsrmsg carries two items as payload:
	(1) a set of \lsadv headers representing the \lsadvs required and, again, (2) the identifier of the sender itself.
	The generation function is
	
\vspace{1.5mm}
	\centerline{$
		\reqID : \pow({\tLSAHDR}) \times \tIP \rightarrow {\tMSG}\,.\spac
	$}
	\item An \lsumsg consists of two fields: (a) an {\tLSDB} and (b) the identifier of the sender.
	The generation function is

\vspace{-2.5mm}
	\centerline{$
		\updID : {\tLSDB} \times \tIP \rightarrow {\tMSG}\,.\spac$}
\end{enumerate}

Sending messages takes time, but should not result in the OSPF process getting stuck. 
Therefore we assume the existence of queues to buffer incoming messages and a sending process, {\QSND}, to handle the sending of messages to other routers (in some sense this is an output queue).

This assumption reflects modern hardware architecture, but bears a problem in modelling.
Messages are unicast to a single destination, groupcast (mulitcast) to a set of destinations or broadcast. 
This information (the sending method and the destinations) is not contained within the OSPF control messages; it needs to be passed onto the sending process {\QSND}.

We define a new (internal) message type that combines the sending method and the destinations with outgoing messages.
This new message type consists of two fields:
(a) the real message that needs to be sent, which is an element of {\tMSG}, and 
(b) a set of destination IPs, which is an element of $\pow(\tIP)$.
As usual we define a generator function
	
	    \vspace{1mm}
	\centerline{$
		\fnsndmsg : {\tMSG} \times \pow(\tIP) \rightarrow {\tMSG}\,.$}
	
	    \vspace{2mm}
\noindent 
The main process {\OSPF} (and its subprocesses) use this message type to pass messages to  the process {\QSND}, which manages the actual sending.

\subsubsection{Update Functions}
We require a set of manipulation functions to change the local data structures such as the LSDB.

When a router discovers a new neighbour it initiates a process to synchronise its own Link State Databases (LSDB) with the LSDB maintained by the neighbour.  
It sends a \emph{database description ({\scshape Dbd}) message} that contains the headers of all the LSAs in its own LSDB to the newly discovered neighbour. 
When the neighbour receives this message it compares all the \lsadv headers against the entries in its own LSDB.
If one is missing or out-dated, it sends a request for that \lsadv. 
The newly discovered neighbour also sends a \dbdmsg to the original router, which triggers a symmetric process.

We generate a new \lsadv for node $\ip$ at time $\t$, using a neighbour structure  $\nbrs$ by 
	\begin{eqnarray*}
		\fnnewLSA : \tIP \times {\tTIME} \times {\tNBRS} & \rightarrow& {\tLSA} \\
		\newLSA\ip\t\nbrs &:=& (\ip, \t, \Pi_1(\nbrs) \})\,,
	\end{eqnarray*}
where $\Pi_i(S):=\{\pi_i(s)\mid s\in S\}$ is the pointwise lifted function $\pi_i$.

To update a node's LSDB with a given set of \lsadvs we use 
	\hypertarget{fn:install}{\begin{eqnarray*}
		\fninstall : {\tLSDB} \times {\tLSDB} & \rightarrow&  {\tLSDB} \\
		\install\lsdb\lsas & :=&\phantom{\cup\ }\{lsa \mid lsa \in \lsdb\ \wedge \not\exists lsa'\in\lsas:\ \hdr{lsa}<\hdr{lsa'}\}\\
															         &&\cup\ \{lsa \mid lsa \in \lsas\ \wedge \not\exists lsa'\in\lsdb:\ \hdr{lsa}\leq\hdr{lsa'}\}\,.
	\end{eqnarray*}}
It is easy to see that this definition is well defined, \ie returns an element of type {\tLSDB}

\noindent The protocol also requires functions to inspect elements of the neighbour list, or to modify it.
\begin{enumerate}
	\item To check whether an entry for $\nip$ exists in $\nbrs$ we use
	\hypertarget{fn:nbrExist}{\begin{eqnarray*}
		\fnnbrExist : {\tNBRS} \times \tIP & \rightarrow& {\tBOOL} \\
		\nbrExist{\nbrs}{\nip}&:\iff& \nip\in\Pi_1(\nbrs)\,.
	\end{eqnarray*}}
	\item To insert a new neighbour structure for node $\nip$ we use the partial function
	\begin{eqnarray*}
		\fnnewNBR : {\tNBRS} \times \tIP & \rightharpoonup& {\tNBRS} \\
		\newNBR{\nbrs}{\nip} & :=& \nbrs \cup \{ (\nip, 0) \} \qquad \text{if } \neg\nbrExist{\nbrs}{\nip}\,.\footnotemark
	\end{eqnarray*}
	
	\vspace{-0mm}
	\footnotetext{For brevity we omit the cases that are undefined.}
    \item We update the inactivity timer of an entry of $\nbrs$ by
	\begin{eqnarray*}
    	\fnsetINACTT : {\tNBRS} \times \tIP \times {\tTIME} &\rightarrow& {\tNBRS} \\
		\setINACTT\nbrs\nip\t & :=& 
		\{ n \mid n \in \nbrs \wedge \pi_1(n) \neq \nip \} \cup \{ (\nip, \t) \}\,.
	\end{eqnarray*}
    \item Given the current time $\t$, nodes need to know which neighbours from $\nbrs$  are inactive. 
	\begin{eqnarray*}
		\fndeadNBRS : {\tNBRS} \times {\tTIME} & \rightarrow& {\tNBRS} \\
		\deadNBRS\nbrs\t & :=& \{ n \mid n \in\nbrs \land \pi_2(n) < \t \}\,.
	\end{eqnarray*}
\end{enumerate}
%

\subsection{The Full Model}
Our simple model, written in the process algebra \tawn, consists of $7$ processes, named {\OSPF}, {\HELLO}, {\DBD}, {\REQ}, {\UPDp}, {\QMSG} and {\QSND}.
\begin{itemize}
\item The main process {\OSPF} reads a single message from the input queue. 
Depending on the type of the message it calls other processes to handle further operations. 
The process also sends \hellomsgs at periodic intervals and maintains the node's LSDB by removing dead neighbours.

\item The process {\HELLO} describes all the actions performed when a \hellomsg is received. 
This includes updating the relevant inactivity timer and, if the \hellomsg is from a previously unknown neighbour, updating the node's {neighbour list}. 
The process also generates a \dbdmsg, which is sent  back to the originator of the \hellomsg, and 
a new \lsadv when required.

\item The process {\DBD} handles incoming \dbdmsgs. 
Sending out a request for those \lsadvs that are described in a \dbdmsg, but do not have a matching partner in the node's LSDB
is among its actions.
\item The process {\REQ} describes the actions following the receipt of an \lsrmsg, such as
 finding the requested \lsadvs in the node's local data, and sending them back to the sender of the message.

\item The process {\UPDp} manages incoming \lsumsgs, including the installation of  up-to-date \lsadvs in the node's LSDB, 
and broadcasting the updated information.

\item The process {\QMSG} models a message queue.
In our model we have two instances.
Whenever a message is received by a node, it is first stored in the first instance, the message input queue. 
If the corresponding node---or more precisely the process {\OSPF}---is able to handle a message it pops the oldest message from the input queue.
 Whenever a message is sent by the process {\OSPF}, it is passed to the second instance of {\QMSG}, which holds the message until the process {\QSND} 
 can handle it.
 
\item The process {\QSND} is responsible for sending outgoing messages that were generated by the other processes described above.
\end{itemize}

\noindent
In the remainder of this  section we present formal specifications of each of these processes and explain them step by step.

\subsubsection{The Basic Routine}
The basic process {\OSPF} (\autoref{pro:simple_basic}) either reads a message from its queue, sends a \hellomsg or removes dead neighbours from its {neighbour list}. 
It maintains four variables: \ip, \nbrs, \lsdb\ and \hellot, in which it stores its own IP address, information about its immediate neighbours, its link state database, and a timestamp stating when the next hello message should be sent.

  \algsetup{linenodelimiter=.,linenosize=\tiny}
  \begin{algorithm}[H]
    {
     \caption{The Basic Routine}
      \label{pro:simple_basic}
      \begin{algorithmic}[1]
\DEFPROCESS{\OSPF}{\ip\comma\nbrs\comma\lsdb\comma\hellot}
	\IFempty									
		\receiveL{\msg}\ .																																	\label{basic_simp_1}
		\COMLINE{depending on the message, the node calls different processes}											\label{basic_simp_2}
		\PAR																																						\label{basic_simp_3}
		\IF[\hellomsg received]{$\msg = \hello{\ips}{\sip}$}																					\label{basic_simp_4}
			\helloL\ips\sip\ip\nbrs\lsdb\hellot																										\label{basic_simp_5}
		\ELSIF[\dbdmsg received]{$\msg = \dbd{\lsahdrs}{\sip}$}																		\label{basic_simp_6}
			\dbdL\lsahdrs\sip\ip\nbrs\lsdb\hellot																									\label{basic_simp_7}
		\ELSIF[\lsrmsg received]{$\msg = \req{\lsahdrs}{\sip}$}																			\label{basic_simp_8}
			\reqL\lsahdrs\sip\ip\nbrs\lsdb\hellot																									\label{basic_simp_9}
		\ELSIF[\lsumsg received]{$\msg = \upd\lsas\sip$}																					\label{basic_simp_10}
			\updL\lsas\sip\ip\nbrs\lsdb\hellot																										\label{basic_simp_11}
		\ENDIFii
		\ENDPAR																																				\label{basic_simp_12}
	\ELSIF[send \hellomsg]{$\now\geq\hellot$}																								\label{basic_simp_13}
		\UPD{\hellot := \now + \hellointvl}																											\label{basic_simp_14}
		\sendL{\sndmsg{\hello{\{\pi_1(n) \mid n \in\nbrs\}}{\ip}}{\emptyset}}\ .														\label{basic_simp_15}
		\ospfL\ip\nbrs\lsdb\hellot																														\label{basic_simp_16}
	\ELSIF[inactive neighbours, which should be removed, exist]{$\deadNBRS{\nbrs}{\now} \neq \emptyset$}	\label{basic_simp_17}
			\UPD{\nbrs := \nbrs - \deadNBRS\nbrs\now}	\COM{remove corresponding neighbour structures}		\label{basic_simp_18}
			\UPD{\lsa := \newLSA\ip\now\nbrs}	\COM{generate a new \lsadv}													\label{basic_simp_19}
			\UPD{\lsdb := \install\lsdb{\{\lsa\}}}		\COM{install it}																			\label{basic_simp_20}
			\sendL{\sndmsg{\upd{\{\lsa\}}{\ip}}{\{\pi_1(n) \mid n\in \nbrs\}}}\ .\COM{and flood it out}						\label{basic_simp_21}
			\ospfL\ip\nbrs\lsdb\hellot																													\label{basic_simp_22}
	\ENDIFii

		\end{algorithmic}
    }
  \end{algorithm}

The message handling is described in Lines \ref{basic_simp_1} to \ref{basic_simp_12}. 
The process first receives a message by \textbf{receive}(\msg); by our specification of the input queue  
(see \autoref{pro:simple_message_queue} below) this will be the message currently in the front of the queue (the oldest message). 
{\OSPF} then checks the type of the message and calls the corresponding process:
in case of a \hellomsg the process {\HELLO} is called, 
in case of a \dbdmsg the process {\DBD}, etc. 

The second part of {\OSPF} (Lines \ref{basic_simp_13} to \ref{basic_simp_16}) ensures that \hellomsgs are sent periodically. 
The parameter $\hellot$ indicates when the next \hellomsg should be sent.
The process checks whether it is time to send the next \hellomsg;
 if so it resets the timer (Line \ref{basic_simp_14}) and  broadcasts a \hellomsg. 
 In fact,  the message is sent to the process {\QSND} (\autoref{pro:simple_sending_queue}), which handles the actual sending. 
 The \hellomsg contains the IP addresses of all the neighbours that the node {\ip} is aware of. 
 After handing over the \hellomsg the node returns to the main process {\OSPF} to handle the next incoming message, or send yet another \hellomsg.

The third part of {\OSPF} (Lines \ref{basic_simp_17} to \ref{basic_simp_22}) collects all of the neighbours from which no activity has been seen for the last {\rtdeadintvl} seconds, removes them from the neighbour structure {\nbrs} and informs the neighbours about these updates.
Line \ref{basic_simp_17} checks whether there exist neighbours from which no activity has been seen for {\rtdeadintvl} seconds, using the function \fndeadNBRS.
If such neighbours exist, the node removes all of the dead IP addresses from the neighbour structure (Line \ref{basic_simp_18}). Then the node generates a new \lsadv with this information (Line \ref{basic_simp_19}), installs this information in its own link state database (Line \ref{basic_simp_20}) and sends out the information to its neighbours (Line \ref{basic_simp_21}).

\subsubsection{Message Processing}
The process {\HELLO} (\autoref{pro:simple_hello_app}) handles the receipt of a \hellomsg. 

\vspace{-1pt}
  \algsetup{linenodelimiter=.,linenosize=\tiny}
  \begin{algorithm}[H]
    {
     \caption{Handling \hellomsgs}
      \label{pro:simple_hello_app}
      \begin{algorithmic}[1]
\DEFPROCESS{\HELLO}{\ips\comma\sip\comma\ip\comma\nbrs\comma\lsdb\comma\hellot}
	\IF[the sender {\sip} is unknown]{$\neg\nbrExist\nbrs\sip$}																				\label{hello_simp_app_1}
		\UPD{\nbrs := \newNBR\nbrs\sip}																												\label{hello_simp_app_2}
		\UPD{\nbrs := \setINACTT\nbrs\sip{\now + \rtdeadintvl}}																				\label{hello_simp_app_3}
		\UPD{\lsa := \newLSA\ip\now\nbrs}\COM{generate new \lsadv}																	\label{hello_simp_app_4}
		\UPD{\lsdb := \install\lsdb{\{\lsa\}}}\COM{ install \lsadv}																				\label{hello_simp_app_5}
		\sendL{\sndmsg{\upd{\{\lsa\}}{\ip}}{\{\pi_1(n) \mid n \in\nbrs\}}}\ . \COM{send \lsadv}								\label{hello_simp_app_6}
		\sendL{\sndmsg{\dbd{\{\hdr{lsa} \mid lsa\in\lsdb\}}{\ip}}{\{\sip\}}}\ .   \COM{send {\sc Dbd} back to \sip}		\label{hello_simp_app_7}
		\ospfL\ip\nbrs\lsdb\hellot 																															\label{hello_simp_app_8}
	\ELSIF[{\sip} is a known neighbour]{$\nbrExist{\nbrs}{\sip}$}																			\label{hello_simp_app_9}
		\UPD{\nbrs := \setINACTT{\nbrs}{\sip}{\now + \rtdeadintvl}}																		\label{hello_simp_app_10}
		\ospfL\ip\nbrs\lsdb\hellot 																															\label{hello_simp_app_11}
	\ENDIFii
		\end{algorithmic}
    }
  \end{algorithm}

\vspace{-1pt}

The first part of the process (Lines \ref{hello_simp_app_1} to \ref{hello_simp_app_8}) is executed
when the originator of the \hellomsg is unknown to the node {\ip}. 
In Line \ref{hello_simp_app_1}, the function \fnnbrExist\ returns a Boolean indicating whether the sender 
{\sip} of the \hellomsg is in the node's {neighbour list} {\nbrs} 
(\ie whether it is known to the node {\ip}). 
In Lines \ref{hello_simp_app_2} to \ref{hello_simp_app_5} the process adds the sender to its neighbour structure (Line \ref{hello_simp_app_2}), resets the inactivity timer for that particular neighbour (Line \ref{hello_simp_app_3}), 
generates an {\lsadv} to reflect the change to the neighbour structure and installs that \lsadv in its link state database. 
Line \ref{hello_simp_app_6} sends out the newly created \lsadv in the form of an \lsumsg, which will be sent to all of the known neighbours by the process \QSND. 
Line \ref{hello_simp_app_7} creates and sends a \dbdmsg, which contains all the headers of the \lsadvs in the link state database.
The message is intended for the newly discovered neighbour \sip, the sender of the {\hellomsg}.

The second part of the process (Lines \ref{hello_simp_app_9} to \ref{hello_simp_app_11}) handles the situation where the sender {\sip} of the \hellomsg is already known to the node. 
All that needs to be done is a reset of the inactivity timer (Line \ref{hello_simp_app_10}), using the function \fnsetINACTT. 
After this it returns to the main process \OSPF.

The process {\DBD} (\autoref{pro:simple_dbd}) handles the receipt of a \dbdmsg. 
It behaves similar to the process {\HELLO} as it checks whether the sender of the incoming message is known to the node or not. 

The first part of the process (Lines \ref{dbd_simp_1} to \ref{dbd_simp_8}) deals with a message from an unknown sender. Lines \ref{dbd_simp_1}--\ref{dbd_simp_7} are identical to the corresponding 
lines of \autoref{pro:simple_hello}: Line \ref{dbd_simp_2} adds the sender to the {neighbour list} \nbrs, Line \ref{dbd_simp_3} resets the neighbour's inactivity timer, Line \ref{dbd_simp_4} generates a new \lsadv to reflect the new state of the {\lsdb} and Line \ref{dbd_simp_5} installs the new \lsadv in the link state database. 
Line \ref{dbd_simp_6} places the newly created LSA in an update message and sends it out to all known neighbours. 
Line \ref{dbd_simp_7} sends a \dbdmsg back to the sender {\sip} of the original message, informing the node about the contents of the node's \lsdb.
After this, the process returns to the start of {\DBD} again, this time the sender is known, as it was added to the neighbour structure {\nbrs} on Line \ref{dbd_simp_2}. This time around the second part of the process is executed.

  \algsetup{linenodelimiter=.,linenosize=\tiny}
  \begin{algorithm}[H]
    {
     \caption{Handling \dbdmsgs}
      \label{pro:simple_dbd}
      \begin{algorithmic}[1]
\DEFPROCESS{\DBD}{\lsahdrs\comma\sip\comma\ip\comma\nbrs\comma\lsdb\comma\hellot}
	\IF[the sender {\sip} is unknown]{$\neg\nbrExist\nbrs\sip$}																				\label{dbd_simp_1}
		\UPD{\nbrs := \newNBR\nbrs\sip}																												\label{dbd_simp_2}
		\UPD{\nbrs := \setINACTT\nbrs\sip{\now + \rtdeadintvl}}																				\label{dbd_simp_3}
		\UPD{\lsa := \newLSA\ip\now\nbrs}\COM{generate new \lsadv}																	\label{dbd_simp_4}
		\UPD{\lsdb := \install\lsdb{\{\lsa\}}}\COM{ install \lsadv}																				\label{dbd_simp_5}
		\sendL{\sndmsg{\upd{\{\lsa\}}{\ip}}{\{\pi_1(n) \mid n\in\nbrs\}}}\ . \COM{send \lsadv}									\label{dbd_simp_6}
		\sendL{\sndmsg{\dbd{\{\hdr{lsa} \mid lsa\in\lsdb\}}{\ip}}{\{\sip\}}}\ .   \COM{send {\sc Dbd} back to \sip}		\label{dbd_simp_7}
		\dbdL\lsahdrs\sip\ip\nbrs\lsdb\hellot																											\label{dbd_simp_8}
	\ELSIF[{\sip} is a known neighbour]{$\nbrExist{\nbrs}{\sip}$}																			\label{dbd_simp_9}
		\sendL{\sndmsg{\req{\{lhdr \mid lhdr \in \lsahdrs \wedge \nexists\, lsa \in \lsdb : lhdr\leq \hdr{lsa}\}}{\ip}}{\{\sip\}}
		}\ . 																																							\label{dbd_simp_10}
		\ospfL\ip\nbrs\lsdb\hellot 																															\label{dbd_simp_11}
	\ENDIFii

		\end{algorithmic}
    }
  \end{algorithm}

\vspace{-10pt}

The second part (Lines \ref{dbd_simp_9} to \ref{dbd_simp_11}) deals with the situation when the sender of the \dbdmsg is already known. 
In this case the \lsadv headers {\lsahdrs} from the \dbdmsg are compared with the node's own link state database {\lsdb}. All the headers that are newer than those in \keyw{lsdb} or that do not yet exist in \keyw{lsdb} are packed into an \lsrmsg and sent back to {\sip} (Line \ref{dbd_simp_10}). 

The process {\REQ} (\autoref{pro:simple_req}) handles the receipt of an \lsrmsg. 
\vspace{-4pt}
  \algsetup{linenodelimiter=.,linenosize=\tiny}
  \begin{algorithm}[H]
    {
     \caption{Handling \lsrmsgs}
      \label{pro:simple_req}
      \begin{algorithmic}[1]
\DEFPROCESS{\REQ}{\lsahdrs\comma\sip\comma\ip\comma\nbrs\comma\lsdb\comma\hellot}
	\IF[the sender {\sip} is unknown]{$\neg\nbrExist\nbrs\sip$}																					\label{req_simp_1}
		\ospfL\ip\nbrs\lsdb\hellot 																																\label{req_simp_2}
	\ELSIF[{\sip} is a known neighbour]{$\nbrExist{\nbrs}{\sip}$}																				\label{req_simp_3}															
		\sendL{\sndmsg{\upd{\{lsa \mid lsa \in \lsdb \wedge \exists\, lhdr\in\lsahdrs : lhdr\leq\hdr{lsa}\}}{\ip}}{\{\sip\}}}\ .					\label{req_simp_4}
		\ospfL\ip\nbrs\lsdb\hellot 																																\label{req_simp_5}														
	\ENDIFii

		\end{algorithmic}
    }
  \end{algorithm}

\vspace{-6pt}
As all other processes it checks whether the sender {\sip} of the message is known. 
If it is unknown (Lines \ref{req_simp_1} and \ref{req_simp_2}) the message is simply ignored and the process returns to {\OSPF}.
If {\sip} is known (Lines~\ref{req_simp_3} to~\ref{req_simp_5}) the node checks the {\lsadv} headers \keyw{lsahdrs} in the received message against the \lsadvs contained in the node's link state database {\lsdb}. 
The LSAs that have headers matching those in the message are packaged into a new \lsumsg and sent back to {\sip}, the sender of the original \dbdmsg.

The process {\UPDp} (\autoref{pro:simple_upd}) handles the receipt of an \lsumsg. 
\vspace{-4pt}
  \algsetup{linenodelimiter=.,linenosize=\tiny}
  \begin{algorithm}[H]
    {
     \caption{Handling \lsumsgs}
      \label{pro:simple_upd}
      \begin{algorithmic}[1]
 \DEFPROCESS{\UPDp}{\lsas\comma\sip\comma\ip\comma\nbrs\comma\lsdb\comma\hellot}
 	\UPD{\lsas := \{lsa \mid lsa \in\lsas \wedge \nexists\, lsa' \in\lsdb : \hdr{lsa} \leq \hdr{lsa'}\}} \COM{more recent \lsadvs}				\label{upd_simp_1}
	\PAR
 	\IF[no new \lsadvs exist]{$\lsas = \emptyset$}																														\label{upd_simp_3}
 		\ospfL\ip\nbrs\lsdb\hellot 																																					\label{upd_simp_4}
 	\ELSIF[new \lsadvs exist]{$\lsas \neq\emptyset$}																													\label{upd_simp_5}
 		\UPD{\lsdb := \install{\lsdb}{\lsas}} \COM{install them}																										\label{upd_simp_6}
 		\sendL{\sndmsg{\upd{\lsas}{\ip}}{\{\pi_1(n) \mid n \in\nbrs\}}}\ .
		\label{upd_simp_7}
 		\ospfL\ip\nbrs\lsdb\hellot 																																					\label{upd_simp_8}
 	\ENDIFii
 	\ENDPAR

		\end{algorithmic}
    }
  \end{algorithm}

In Line \ref{upd_simp_1}, the process finds those {\lsadvs} in the payload of the message that are either more recent than those in the link state database \lsdb, or that do not yet exist in \lsdb; these are stored in the variable {\lsas}.
In case no such \lsadv exists, \ie the set {\lsas} is empty, the process does not perform any action and returns to the main process {\OSPF} (Lines \ref{upd_simp_3} and \ref{upd_simp_4}).
In case the incoming message contains new information (Line~\ref{upd_simp_5}), \ie $\lsas$ is not empty, 
the process {\fninstall}s the newly discovered \lsadvs into its own database (Line \ref{upd_simp_6}).
Since the node has received new \lsadvs, it forwards them to its neighbours:
 Line \ref{upd_simp_7} packages the set {\lsas} into an update message and sends them to the neighbours. 
 Lastly, the process returns to {\OSPF}.

\subsubsection{Queues}

The process {\QMSG} (\autoref{pro:simple_message_queue}), which is identical to the message queue presented in~\cite{TR13}, 
handles the buffering of messages that are to be sent or received. By this we guarantee that no message is lost, 
regardless of which state the main process {\OSPF} is in--- {\OSPF} can
reach a state, such as {\HELLO}, {\DBD}, {\REQ} or {\UPDp}, in which
it is not ready to perform a receive action. 

  \algsetup{linenodelimiter=.,linenosize=\tiny}
  \begin{algorithm}[H]
    {
     \caption{Message Queue}
      \label{pro:simple_message_queue}
      \begin{algorithmic}[1]

\DEFPROCESS{\QMSG}{\msgs}
	\IFempty
		\COMLINE{store incoming message at the end of \msgs}										\label{queue_simp_1}
		\receiveL{\msg}\ . 			
		\qmsgP{\append{\msg}{\msgs}}																				\label{queue_simp_2}
	\ELSIF[the queue is not empty]{$\msgs\not=[\,]$}														\label{queue_simp_3}
		\PAR
		\COMLINE{pop top message and send it to another sequential process}				\label{queue_simp_4}
		\sendL{\head{\msgs}}\ .\ \qmsgP{\tail{\msgs}}															\label{queue_simp_5}
		\COMLINE{or receive and store an incoming message}											\label{queue_simp_7}
		\STATE $+$\,  \receive{\msg}\ . \qmsgP{\append{\msg}{\msgs}}                  		    \label{queue_simp_8}
		\ENDPAR
	\ENDIFii
		\end{algorithmic}
    }
  \end{algorithm}

\vspace{-4mm}

The process {\QMSG} runs in parallel with {\OSPF} or any other process that might be called and 
maintains a queue (list) {\msgs} of messages received. 
In fact, every node maintains two {\QMSG} processes, one handling incoming messages and the other handling outgoing messages.

It is always ready to receive yet another new message {\msg} and append it to \msgs.

The process {\QSND} (\autoref{pro:simple_sending_queue}) handles the sending of messages. Similar to {\QMSG}, a copy of {\QSND} runs in parallel with {\OSPF} or any other process that is called. 
It receives the next message to be sent in Line~\ref{qsnd_simp_1}. 
Depending on the type of the message, the process either broadcasts the message (Lines \ref{qsnd_simp_4} and~\ref{qsnd_simp_5}), or 
groupcasts it to the destinations specified in the message received (Lines \ref{qsnd_simp_6} and \ref{qsnd_simp_7}). 

  \algsetup{linenodelimiter=.,linenosize=\tiny}
  \begin{algorithm}[H]
    {
     \caption{Sending Process}
      \label{pro:simple_sending_queue}
      \begin{algorithmic}[1]
\DEFPROCESS{\QSND}{}
	\IFempty
		\receiveL{\msg}\ .																														\label{qsnd_simp_1}
		\IF[get content of \msg]{$\msg = \sndmsg{\realmsg}{\dips}$}														\label{qsnd_simp_2}
			\PAR																																		\label{qsnd_simp_3}
				\IF[broadcast \hellomsg]{$\realmsg = \hello\ips\sip$}															\label{qsnd_simp_4}
					\broadcast{\realmsg}\ . \qsndP																						\label{qsnd_simp_5}
				\ELSIF[groupcast other messages]{$\bigwedge_{\ips,\sip} \realmsg \neq \hello\ips\sip$}		\label{qsnd_simp_6}
					\groupcast{\dips}{\realmsg}\ . \qsndP																				\label{qsnd_simp_7}
				\ENDIFii
			\ENDPAR																																\label{qsnd_simp_8}
		\ENDIFii
	\ENDIFii
		\end{algorithmic}
    }
  \end{algorithm}

\subsubsection{Initial State\label{sec:init_simp}}
We need to define an initial state to finish our specification. The initial expression of the network is an encapsulated parallel composition of node expressions $\ip{:}P{:}R$, where the finite number of nodes and the range $R$ of each node expression is left unspecified. Each node in the parallel composition must have a unique IP address \dval{ip}. The initial process $P$ for each \dval{ip} is given by the following expression: 
\[
(\xi_0, \qsndP) \parl (\zeta, {\QMSG}(\msgs)) \parl (\chi, {\OSPF}(\ip,\nbrs,\lsdb,\hellot)) \parl (\theta,{\QMSG}(\msgs))
\]
with $\zeta(\msgs) = [\,]$, $\chi(\ip) = \dval{ip} \land \chi(\nbrs) = \chi(\lsdb) = \emptyset \land \chi(\hellot) = 0$ and $\theta(\msgs) = [\,]$.
As {\QSND} does not maintain any variable its valuation is empty.

Initially both the outgoing and incoming message queues are empty.
The process {\OSPF} knows its own IP address, is not aware of any neighbour, has an empty \keyw{LSDB} and the hello timer is set to zero, so it will send a \hellomsg as soon as possible.

A node will not lose any messages received even if they arrive while the process {\OSPF} is busy. 
This is because the rightmost {\QMSG} process is running in parallel and can receive and store any incoming message. 
As soon as {\OSPF} can handle another message, it synchronises with the rightmost {\QMSG}.
The leftmost {\QMSG} performs a similar role for outgoing messages; the processes passes on to {\QSND}, which sends the messages to the incoming {\QMSG} process of other nodes.

\newpage
\section{The Detailed AWN Model\label{app:modelI}}
In this section we provide an in-depth description of our detailed model of OSPF, which we sketched in \autoref{sec:modelI}. 
To simplify the description of this model, many aspects are described by how they differ from the simpler model (\autoref{sec:modelII} and \autoref{app:modelII}). 

\subsection{Data Structure}
The types {\tLSA} and {\tLSDB} remain unchanged (see \autoref{app:modelII} for their definition), but the data structures dealing with neighbours carry a lot more information and as such they need to be altered. 

There are two main reasons why these structures need to be more complex. 
First, we assume that messages can now be lost, therefore the protocol has to implement message retransmission as well as message acknowledgments. 
Secondly, it is no longer necessary for a node to form an adjacency with each of its neighbours.

\subsubsection{Altered Data Structure Relating to Neighbours}

Next to the node's IP address and time stamp, a neighbour structure now contains seven more fields: 
one for neighbour states, two to enable  reliable Database Description exchange, two for \lsrmsg retransmission, and two for \lsumsg retransmission. 

We first introduce the concept of a \emph{neighbour state} ({\tNS}). It is a string that represents the six phases of OSPF's adjacency establishment.\\[2mm]
\centerline{${\tNS} := \str{Init} \>\> \lvert \>\> \str{2-Way} \>\> \lvert \>\> \str{ExStart} \>\> \lvert \>\> \str{Exchange} \>\> \lvert \>\> \str{Loading} \>\> \lvert \>\> \str{Full}$}\\[2mm]
\noindent 
A node associates a {neighbour state} with each of its neighbours. 
The state describes whether the two nodes {\ip} and {\sip} should exchange \lsadvs and if so, it describes the stage of LSDB synchronisation.
We explain the meaning of the strings in the following:
\begin{description}
	\item[Init:]
	A node {\ip} has recently received a \hellomsg from a neighbour. However, bidirectional communication is 
		not confirmed yet---the node's own IP address is not part of the \hellomsg. 
	\item[2-Way:]
		In this state communication between the two nodes  has been confirmed to be bidirectional; 
		 only neighbours whose adjacency is not needed will enter \str{2-Way}.
	\item[ExStart:]
		This is the first step in the adjacency-establishment procedure. It establishes the master/slave relationship between the nodes.
	\item [Exchange:]
		A node in this state exchanges its link state database, using \dbdmsgs.
	\item[Loading:] The nodes have exchanged descriptions of their LSDBs and now compare the information with their local data. 
		Missing information is requested by \lsrmsgs.
	\item [Full:] This state indicates full adjacency, \ie the nodes have exchanged all information and their LSDBs are synchronised.
\end{description}
\noindent We define a strict order on {\tNS}. It reflects the progress in the adjacency-establishment procedure: the greater the value the further the procedure has evolved.\\[2mm]
\centerline{$
	\str{Init} < \str{2-Way} < \str{ExStart} < \str{Exchange} < \str{Loading} < \str{Full}
$}\\[2mm]
Following the RFC,  we call neighbours that are in state \str{ExStart} or greater \emph{adjacencies}.

Before defining the altered neighbour structure we need to introduce the concept of sequence numbers, which were not present in the simple model.
For that, we introduce the type {\tSQN}, which we assume to be isomorphic to \NN.
Database Description Sequence Numbers (DDSQNs), used in \dbdmsgs, ensure that both nodes correctly communicate.
Every neighbour structure stores one sequence number, called a DD sequence number in the RFC.
As usual, sequence numbers indicate the freshness of a \dbdmsg.
If a node receives a \dbdmsg  with a larger than expected sequence number the adjacency-establishment procedure needs to be restarted

\noindent We can now formally describe a \emph{neighbour structure}; it is a tuple with nine components:
\begin{enumerate}
	\item The IP address of a neighbour, which is an element of {\tIP} (as in the simple model).
	\item A neighbour state, an element of {\tNS}.
	\item  A timestamp of type \tTIME, which indicates when the neighbour is considered inactive or dead---unless another message is received from this neighbour before that time (as in the simple model).
	\item A DD sequence number, which is an element of {\tSQN}.
	\item A timestamp of type $\tTIME$ indicating when the DD timer will fire, which triggers resending of a \dbdmsg.
	\item A link state request list\footnote{It is called list, but the order does not matter}, 
	an element of $\pow({\tLSAHDR})$, containing the headers of all the \lsadvs that need to be requested from that neighbour.
	\item A timestamp indicating when the next \lsrmsg will be sent;
	\item A link state retransmission list, an element of {\tLSDB} that contains all the \lsadvs that have been sent to the neighbour and are yet to be acknowledged.
	\item A timestamp indicating when the link state retransmission timer will fire.
\end{enumerate}

\noindent We denote the type of neighbour structures by {\tNBR} and define a generation function:
\[
	\keyw{(\_,\_,\_,\_,\_,\_,\_,\_,\_)} : {\tIP} \times {\tNS} \times \tTIME \times {\tSQN} \times \tTIME \times \pow(\tLSAHDR) \times \tTIME \times {\tLSDB} \times \tTIME \rightarrow {\tNBR}
	\]
\noindent Identically to the simple model, a \emph{neighbour list} is a set of neighbour structures, where each IP address has at most one entry. 
\[
	{\tNBRS} := \{ nbrs \>\> \lvert \>\> nbrs \in \pow({\tNBR}) \>\> \land \>\> \forall nbr_1, nbr_2 \in nbrs : nbr_1 \neq nbr_2 \Rightarrow \pi_1(nbr_1) \neq \pi_1(nbr_2) \}\,.
\]
Of course they now contain the more complicated neighbour structures.

\subsubsection{Messages}
We also have to modify messages: \dbdmsgs contain more information; \lsrmsgs carry only one \lsadv header rather than a set of them; and \ackmsgs are introduced as a new message type. In detail we have the following:

\begin{enumerate}
	\item A \hellomsg remains unchanged and contains two entries: (a) a set of IP addresses of all neighbours 
		discovered by the sender and (b) the identifier of the sender itself. The generation function is
		
		\vspace{-2mm}
		\centerline{$\helloID : \pow(\tIP) \times \tIP \rightarrow {\tMSG}\,.$\spac}
	\item A \dbdmsg consists of four fields:
		(a) a set of \lsadv headers describing the LSDB of the sender (as in the simple model);
		(b) a DD sequence number for keeping the stream of messages reliable and ordered;
 		(c) an init bit, an element of {\tBOOL}, indicating whether the message is the first in a sequence when set to true; and
		(d)  the identifier of the sender itself (as in the simple model).
		The generation function is
		
		\vspace{-2mm}
		\centerline{$\dbdID : \pow({\tLSAHDR}) \times {\tSQN} \times {\tBOOL} \times {\tIP} \rightarrow {\tMSG}\,.$}
		
	\item An \lsrmsg still consists of 2 fields, but the first is changed:
		(a) a single \lsadv header of type {\tLSAHDR}; and, again, 
		(b) the identifier of the sender itself. The generation function is
		
		\vspace{2mm}
		\centerline{$\reqID : {\tLSAHDR} \times {\tIP} \rightarrow {\tMSG}\,.$}
		
		\vspace{2mm}
	\item An \lsumsg remains unchanged and contains (1) an {\tLSDB} and (2) the identifier of the sender.
		It is generated by $\updID : {\tLSDB} \times \tIP \rightarrow {\tMSG}$.
	\item A \emph{Link State Acknowledgement \ackmsg} consists of 2 fields:
		(a) a set of \lsadv headers, an element of $\pow({\tLSAHDR})$; and, again, 
		(b) the identifier of the sender itself. It is generated by
		
		\vspace{2mm}
		\centerline{$\ackID : \pow({\tLSAHDR}) \times {\tIP} \rightarrow {\tMSG}\,.$}
\end{enumerate}

\subsubsection{Update Functions}
As we take the adjacency-establishment procedure into account we need further functions.
We also have to alter functions that take the new data structure into account. 
We also introduce auxiliary functions to better structure both the data structure and the processes.

As discussed in \autoref{sec:modelI} we assume the existence of a predicate $\adj\ip{\ip'}$, 
which evaluates to true if the nodes {\ip} and $\ip'$ are supposed to form an adjacency.\\[2mm]
		\centerline{$\begin{array}{rcl}
	\fnadj : {\tIP} \times {\tIP} & \rightarrow& {\tBOOL}\\
	\adj\ip{\ip'} &:\iff& \text{adjacency should be established between } \ip \text{ and } \ip'
\end{array}$}\\[2mm]
We further assume that $\adj{\ip}{\ip'}=\adj{\ip'}{\ip}$.

\noindent We first consider all the functions that concern LSDBs.
\begin{enumerate}
	\item The generation of \lsadvs now takes the neighbour state into account:
	
	    \vspace{2mm}
		\centerline{$\begin{array}{rcl}
		\fnnewLSA : {\tIP} \times \tTIME \times {\tNBRS} & \rightarrow& {\tLSA} \\
		\newLSA\ip\t\nbrs & :=& (\ip, \t, \{\pi_1(n) \>\> \lvert \>\> n \in \nbrs \land \pi_2(n) \geq \str{2-Way} \})
\end{array}$}
	
	    \vspace{2mm}
	   
      \item The function \hyperlink{fn:install}{\fninstall}, which updates a node's LSDB, remains unchanged.
	\item To check whether the LSDB contains up to date information for a given \lsadv header, we use
	
	    \vspace{2mm}
		\centerline{$\begin{array}{rcl}
		\fnlsaExist : {\tLSDB} \times {\tLSAHDR} & \rightarrow& {\tBOOL} \\
		\lsaExist\lsdb\lsahdr &:\iff& \exists\, lsa \in \lsdb : \lsahdr \leq \hdr{lsa}
\end{array}$}
	
	    \vspace{2mm}
	\item 		A node has to find an entry in its LSDB, given an \lsadv header.
	
	    \vspace{2mm}
		\centerline{$\begin{array}{rcl}
			\fngetLSA : {\tLSDB} \times {\tLSAHDR} & \rightharpoonup &{\tLSA} \\
			\getLSA\lsdb\lsahdr & :=&
			lsa\qquad  \text{if } \exists\, lsa \in \lsdb \wedge \pi_1(lsa) = \pi_1(\lsahdr) 
\end{array}$}
\end{enumerate}

\medskip
\noindent As before, the protocol also requires functions to inspect elements of the neighbour list, or to modify it.

\begin{enumerate}[resume]
	\item The function \hyperlink{fn:nbrExist}{\fnnbrExist}, which checks whether a certain entry exists, remains unchanged.
	\item The function that inserts a new neighbour structure needs alteration to reflect the new structure.
	
	    \vspace{2mm}
		\centerline{$\begin{array}{rcl}
		\ \ \fnnewNBR : {\tNBRS} \times {\tIP} &\rightharpoonup& {\tNBRS} \\
		\newNBR{\nbrs}{\nip} & :=& \nbrs \cup \{ (\nip, \str{Init}, 0, 0, 0, \emptyset, 0, \emptyset, 0) \}\qquad \text{if } \neg\nbrExist{\nbrs}{\nip}
	\end{array}$}
	
	    \vspace{2mm}
\item Given an IP address we need to find the corresponding neighbour structure in the neighbour list.
	
	    \vspace{2mm}
		\centerline{$\begin{array}{rcl}
		\fngetNBR : {\tNBRS} \times {\tIP} & \rightharpoonup& {\tNBR} \\
		\getNBR\nbrs\nip & := &n \qquad \text{if }\exists\, n\in \nbrs : \pi_1(n) = \nip  
		\end{array}$}
	
	    \vspace{2mm}
	The function $\getNBR\nbrs\nip$ is defined iff $\nbrExist\nbrs\nip$. By the type definition of {\tNBRS} the entry is unique.
\end{enumerate}
\pagebreak
\noindent We also define functions that return individual components of the corresponding neighbour structure.
\begin{enumerate}[resume]	
	\item Return the neighbour state.
	
		\vspace{1mm}
		\centerline{$\begin{array}{rcl}
		\fngetNS : {\tNBRS} \times {\tIP} & \rightharpoonup& {\tNS} \\
		\getNS\nbrs\nip & := &
			\pi_2(\getNBR\nbrs\nip) \qquad \text{if }\nbrExist\nbrs\nip
	\end{array}$}
	
			\vspace{2mm}
	\item Return the DD sequence number.
	
		\vspace{1mm}
		\centerline{$\begin{array}{rcl}
		\fngetDDSQN : {\tNBRS} \times {\tIP} & \rightharpoonup& {\tSQN} \\
		\getDDSQN\nbrs\nip& := &\pi_4(\getNBR\nbrs\nip) \qquad \text{if }\nbrExist\nbrs\nip
	\end{array}$}
	
			\vspace{2mm}
	\item Return the link state request list.
	
		\vspace{1mm}
		\centerline{$\begin{array}{rcl}
		\fngetREQS : {\tNBRS} \times {\tIP} & \rightharpoonup& \pow(\tLSAHDR) \\
		\getREQS\nbrs\nip & :=&
			\pi_6(\getNBR\nbrs\nip) \qquad \text{if }\nbrExist\nbrs\nip
	\end{array}$}
	
			\vspace{2mm}
	\item Return the link state retransmission list.
	
		\vspace{1mm}
		\centerline{$\begin{array}{rcl}
		\fngetRXMTS : {\tNBRS} \times {\tIP} & \rightharpoonup& {\tLSDB} \\
		\getRXMTS\nbrs\nip & :=& \pi_8(\getNBR\nbrs\nip)  \qquad \text{if }\nbrExist\nbrs\nip
	\end{array}$}
	
			\vspace{3mm}
\end{enumerate}
\noindent Dual to reading values from a neighbour structure we need to insert and alter entries.
\begin{enumerate}[resume]
	\item Initialise a neighbour state to $ns$ for a given IP address $nip$. 
		Both the link state request list and the set of link state retransmissions are set to empty.
	
		\vspace{2mm}
		\centerline{\hspace{18pt} $\begin{array}{rcl}
			\fninitNBR : {\tNBRS} \times {\tIP} \times {\tNS} & \rightarrow& {\tNBRS} \\
			\initNBR\nbrs\nip\ns & := &\{ n \mid n \in\nbrs \wedge \pi_1(n) \neq \nip \} \\
			&& \hspace{-2cm}\cup\ \{ (\pi_1(n), \ns, \pi_3(n), \pi_4(n), \pi_5(n), [\ ], \pi_7(n), \emptyset, \pi_9(n)) \mid n \in nbrs \land \pi_1(n) = \nip \}
	\end{array}$}
	
			\vspace{2mm}
	\item Update the neighbour state of a given IP address.
	
		\vspace{2mm}
		\centerline{$\begin{array}{rcl}
			\fnsetNS : {\tNBRS} \times {\tIP} \times {\tNS} &\rightarrow& {\tNBRS} \\
			\setNS\nbrs\nip\ns & :=& \{ n \mid n \in \nbrs \wedge \pi_1(n) \neq \nip \} \\
			&& \hspace{-3.6mm}\cup\ \{(\pi_1(n), \ns, \pi_3(n),\dots,\pi_9(n)) \mid n \in \nbrs \land \pi_1(n) = \nip \}
	\end{array}$}
	
			\vspace{2mm}
		In case no entry exists for $\nip$, the neighbour structure remains unchanged. The same applies to most of the functions defined in the remainder of the section.
	\item Update the activity timer.
	
		\vspace{2mm}
		\centerline{\hspace{18pt} $\begin{array}{rcl}
   	 		\fnsetINACTT : {\tNBRS} \times \tIP \times {\tTIME} &\rightarrow& {\tNBRS} \\
			\setINACTT\nbrs\nip\t & :=& 
			\{ n \mid n \in \nbrs \wedge \pi_1(n) \neq \nip \}\\
			&& \hspace{-3.6mm}\cup\ \{(\pi_1(n), \pi_2(n), \t, \pi_4(n),\dots, \pi_9(n)) \mid n \in \nbrs \land \pi_1(n) = \nip \}
	\end{array}$}
	
			\vspace{2mm}
	\item Update the DD sequence number. 
		We have a function to set the sequence number to an arbitrary value and one for incrementation.
	
		\vspace{2mm}
		\centerline{\hspace{18pt} $\begin{array}{rcl}
    			\fnsetDDSQN : {\tNBRS} \times {\tIP} \times {\tSQN} &\rightarrow& {\tNBRS} \\
			\setDDSQN\nbrs\nip\sqn & :=& 
			\{ n \mid n \in \nbrs \wedge \pi_1(n) \neq \nip \}\\
			&& \hspace{-2cm}\cup\ \{(\pi_1(n),\dots,\pi_3(n), \sqn, \pi_5(n),\dots,\pi_9(n)) \mid n \in \nbrs \land \pi_1(n) = \nip \}\\
		\fnincDDSQN : {\tNBRS} \times {\tIP} &\rightarrow&{\tNBRS} \\
		\incDDSQN\nbrs\nip & := & \{ n \mid n \in \nbrs \wedge \pi_1(n) \neq \nip \}\\
			&& \hspace{-2cm}\cup\ \{(\pi_1(n),\dots,\pi_3(n), \pi_4(n){+}1, \pi_5(n),\dots,\pi_9(n)) \mid n \in \nbrs \land \pi_1(n) = \nip \}
	\end{array}$}
	
			\vspace{2mm}
	\item Update the DD timer.
	
		\vspace{2mm}
		\centerline{\hspace{18pt}$\begin{array}{rcl}
			\fnsetDDT : {\tNBRS} \times {\tIP} \times \tTIME &\rightarrow& {\tNBRS}\\
			\setDDT\nbrs\nip\t & :=& \{ n \mid n \in \nbrs \wedge \pi_1(n) \neq \nip \}\\
			&& \hspace{-3.6mm}\cup\ \{(\pi_1(n),\dots,\pi_4(n), \t, \pi_6(n),\dots,\pi_9(n)) \mid n \in \nbrs \land \pi_1(n) = \nip \}
	\end{array}$}
	\item We update the link state request list in several ways. First we can replace that list with a given set.
	
		\vspace{2mm}
		\centerline{\hspace{18pt}$\begin{array}{rcl}
			\fnsetREQS : {\tNBRS} \times {\tIP} \times \pow(\tLSAHDR)&\rightarrow& {\tNBRS}\\
			\setREQS\nbrs\nip\reqs & :=& \{ n \mid n \in \nbrs \wedge \pi_1(n) \neq \nip \}\\
			&& \hspace{-2.5cm}\cup\ \{(\pi_1(n),\dots,\pi_5(n), \reqs, \pi_7(n),\dots,\pi_9(n)) \mid n \in \nbrs \land \pi_1(n) = \nip \}
	\end{array}$}
	
			\vspace{2mm}
		\noindent We also want to update the list by removing \lsadv headers that are dominated by elements in {\lsdb}.
	
		\vspace{2mm}
		\centerline{$\begin{array}{rcl}
			\fncleanREQS : {\tNBRS} \times {\tIP} \times {\tLSDB} \hspace{-2mm}&\rightharpoonup&\hspace{-2mm} {\tNBRS}\\
			\cleanREQS\nbrs\nip\lsdb \hspace{-2mm}&:=&\hspace{-2mm}\setREQS\nbrs\nip{reqs}\qquad \text{if }\nbrExist\nbrs\nip\,,
	\end{array}$}
	
			\vspace{2mm}
		where $reqs = \{ lhdr \in \getREQS\nbrs\nip\mid \nexists\, lsa \in \lsdb: lhdr \leq \hdr{lsa}\}$.\\
		Moreover, when receiving a \dbdmsg, the protocol wants to add elements to the link state request list; this is combined with 
		cleaning up the list, \ie removing elements that are dominated.
	
		\vspace{2mm}
		\centerline{$\begin{array}{rcl}
			\fnaddREQS : {\tNBRS} \times {\tIP} \times {\tLSDB} \times \pow({\tLSAHDR}) &\rightharpoonup& {\tNBRS}\\
			\addREQS\nbrs\nip\lsdb\lsahdrs&:=&\cleanREQS{nbrs'}\nip\lsdb\,,
	\end{array}$}
	
			\vspace{2mm}
		where $nbrs' = \setREQS\nbrs\nip{\getREQS{\nbrs}{\nip} \cup \lsahdrs}$. \\\noindent
		The set $nbrs'$ is the neighbour list {\nbrs} where  {\nip}'s request list is extended by all the \lsadv headers $\lsahdrs$.
		The function {\fncleanREQS} removes all those headers that are dominated by \lsdb.
	\item Update the link state request timer.
	
			\vspace{2mm}
		\centerline{\hspace{18pt}$\begin{array}{rcl}
			\fnsetREQT : {\tNBRS} \times {\tIP} \times \tTIME &\rightarrow& {\tNBRS}\\
			\setREQT\nbrs\nip\t& :=& \{ n \mid n \in \nbrs \wedge \pi_1(n) \neq \nip \}\\
			&& \hspace{-3.6mm}\cup\ \{(\pi_1(n),\dots,\pi_6(n), \t, \pi_8(n),\pi_9(n)) \mid n \in \nbrs \land \pi_1(n) = \nip \}
	\end{array}$}
	
			\vspace{2mm}
	\item An update to the link state retransmission list, an element of type {\tLSDB}, happens in different ways.
		Firstly, we can set the retransmission list directly.
	
		\vspace{2mm}
		\centerline{$\begin{array}{rcl}
			\fnsetRXMTS :{\tNBRS} \times {\tIP} \times {\tLSDB} &\rightarrow& {\tNBRS}\\
			\setRXMTS\nbrs\nip\lsas& :=& \{ n \mid n \in \nbrs \wedge \pi_1(n) \neq \nip \}\\
			&& \hspace{-3.6mm}\cup\ \{(\pi_1(n),\dots,\pi_7(n), \lsas,\pi_9(n)) \mid n \in \nbrs \land \pi_1(n) = \nip \}
	\end{array}$}
	
			\vspace{2mm}
		Secondly, given a set of \lsadv headers we remove corresponding \lsadvs from the retransmission list.
	
			\vspace{2mm}
		\centerline{\hspace{18pt}$\begin{array}{rcl}
			\fncleanRXMTS : {\tNBRS} \times {\tIP} \times \pow({\tLSAHDR}) \hspace{-2mm}&\rightharpoonup&\hspace{-2mm} {\tNBRS} \\
			\cleanRXMTS\nbrs\nip\lsahdrs \hspace{-2mm}&:=&\hspace{-2mm} 
			\setRXMTS\nbrs\nip{rxmts} \quad\text{if }\nbrExist\nbrs\nip
	\end{array}$}
	
			\vspace{2mm}
		where $rxmts =\{lsa\mid lsa\in\getRXMTS\nbrs\nip \wedge \nexists\, lhdr\in\lsahdrs : \hdr{lsa}\leq lhdr\}$.\\
		For a given LSDB the update needs to change all the neighbour structures with a neighbour state that is equal to or greater than \str{Exchange}, 
		     only keeping the most recent version of each \lsadv.
	
		\vspace{2mm}
		\centerline{$\begin{array}{rcl}
			\fnupdRXMTS : {\tNBRS} \times {\tLSDB} &\rightarrow& {\tNBRS} \\
			\updRXMTS\nbrs\lsas & :=& \{\updRX{n}{\lsas} \mid n \in\nbrs \wedge \pi_2(n) \geq \str{Exchange} \} \\
													&& \hspace{-3.6mm}\cup\ \{n \mid n \in\nbrs \wedge \pi_2(n) < \str{Exchange} \}\, 
	\end{array}$}
	
			\vspace{2mm}
		where the function {\fnupdRX} installs the \lsadvs in the given neighbour structure.
	
			\vspace{2mm}
		\centerline{$\begin{array}{rcl}
			\fnupdRX : {\tNBR} \times {\tLSDB} &\rightarrow& {\tNBR}\\
			\updRX{\ns}{\lsas} &:=& (\pi_1(\ns), ..., \pi_7(\ns), \install{\pi_8(\ns)}{\lsas}, \pi_9(\ns))
	\end{array}$}
	
			\vspace{2mm}
	\item Update the link state retransmission list timer.
	
		\vspace{1mm}
		\centerline{$\begin{array}{rcl}
			\fnsetRXMTT : {\tNBRS} \times {\tIP} \times \tTIME &\rightarrow& {\tNBRS}\\
			\setRXMTT\nbrs\nip\t& :=& \{ n \mid n \in \nbrs \wedge \pi_1(n) \neq \nip \}\\
			&& \hspace{-3.6mm}\cup\ \{(\pi_1(n),\dots,\pi_8(n), \t) \mid n \in \nbrs \land \pi_1(n) = \nip \}
	\end{array}$}
\end{enumerate}
\pagebreak
\noindent The following five functions distil information out of a neighbour structure, other than individual values. 
\begin{enumerate}[resume]	
    \item The function {\fndeadNBRS} keeps the same functionality as in the simple model, namely identifying inactive neighbours. 
    		However, the function needs adaptation to the new neighbour structure.
	
			\vspace{2mm}
		\centerline{$\begin{array}{rcl}
			\fndeadNBRS : {\tNBRS} \times {\tTIME} & \rightarrow& \tNBRS \\
			\deadNBRS\nbrs\t & :=& \{ n \mid n \in\nbrs \land \pi_3(n) < \t \}
	\end{array}$}
	
			\vspace{2mm}
\end{enumerate}
\noindent For the following selections, we use a function $\fnsel: \pow(\tIP)\rightharpoonup\tIP$ that (deterministically) chooses an element 
from a set of IP addresses.
\begin{enumerate}[resume]
	\item Given a time $\t$, return the IP of a neighbour whose DD timer has fired and has a neighbour state which is \str{ExStart} or is the slave in an ongoing exchange process.
	
			\vspace{2mm}
		\centerline{$\begin{array}{rcl}
		\fnddNIP : {\tNBRS} \times \tTIME \times {\tIP} & \rightharpoonup& {\tIP} \\
		\ddNIP\nbrs\t\ip& :=& \sel{ddnips} \qquad \text{if } ddnips \neq \emptyset\,,
	\end{array}$}
	
			\vspace{2mm}
	where $ddnips = \{ \pi_1(n) \mid n \in \nbrs \land \pi_5(n) < \t \wedge((\pi_2(n) = \str{ExStart}) \lor (\pi_2(n) = \str{Exchange} \land \pi_1(n) \leq ip)) \}.$
	Note that this function requires a total order on \tIP, which we assume to exist. We use this order in the establishment of the master/slave 
	relationship between two nodes: the one with the larger IP address will be the master.
\item Given a neighbour structure and a time, the protocol picks a neighbour whose link state request timer has fired to start a communication.
	
			\vspace{2mm}
		\centerline{$\begin{array}{rcl}
		\fnreqNIP : {\tNBRS} \times \tTIME & \rightharpoonup& {\tIP} \\
		\reqNIP\nbrs\t & :=& \sel{reqnips} \qquad \text{if } reqnips \neq \emptyset\,,
	\end{array}$}
	
			\vspace{2mm}
	where $reqnips = \{ \pi_1(n)\mid n \in\nbrs \land \pi_7(n) < \t \land \pi_6(n) \neq \emptyset \}$.\\
	Similarly to this, the protocol also picks a node whose link state retransmission timer has fired, and that has a non-empty link state retransmission list.
	
			\vspace{2mm}
		\centerline{$\begin{array}{rcl}
		\fnrxmtNIP : {\tNBRS} \times \tTIME & \rightharpoonup& {\tIP} \\
		\rxmtNIP\nbrs\t &:=&\sel{rxmtnips} \qquad \text{if } rxmtnips \neq \emptyset\,, 
	\end{array}$}
	
			\vspace{2mm}
	where $rxmtnips = \{ \pi_1(n) \mid n \in \nbrs \land \pi_9(n) < \t \land \pi_8(n) \neq \emptyset \}$.
	\item Get the IPs of all of the neighbours whose neighbour state $\geq \str{Exchange}$.
	
			\vspace{2mm}
		\centerline{$\begin{array}{rcl}
		\fnfloodNIPS : {\tNBRS} & \rightarrow& \pow({\tIP}) \\
		\floodNIPS\nbrs & :=& \{ \pi_1(n) \mid n \in \nbrs \land \pi_2(n) \geq \str{Exchange} \}
	\end{array}$}
	
			\vspace{2mm}
\end{enumerate}

\noindent OSPF only sends messages to those neighbours that have a status equal to or greater than \str{ExStart}. 
\begin{enumerate}[resume]
	\item We define a function that generates a  \dbdmsg intended for node $\nip$.
	
			\vspace{2mm}
		\centerline{$\begin{array}{rcl}
		&&\fngenDBD : {\tNBRS} \times {\tLSDB} \times {\tIP} \times {\tIP} \rightharpoonup {\tMSG}\\
		&&\genDBD\nbrs\lsdb\nip\ip :=\left\{ \begin{array}{ll}
				\dbdDetailed{\{\hdr{lsa} \mid lsa \in \lsdb \}}{\getDDSQN\nbrs\nip}{\true}{\ip}\\
						\hspace{19mm} \text{if }\nbrExist\nbrs\nip \land \pi_2(n) = \str{ExStart} \\
				\dbdDetailed{\emptyset}{\getDDSQN\nbrs\nip}{\false}{\ip}\\
						\hspace{19mm} \text{if }\nbrExist\nbrs\nip \land\pi_2(n) \geq \str{Exchanging}
\end{array}\right.
	\end{array}$}
	
	\vspace{1mm}
\end{enumerate}
\subsection{The Full Model}

The detailed model consists of the 9 processes {\OSPF}, {\HELLO}, {\DBD}, \keyw{SNMIS}, \keyw{REQ}, {\UPDp}, \keyw{ACK}, {\QMSG} and {\QSND}.
The main functionality of the processes already present in the simple model do not change. For completeness, we list a short summary nevertheless.
\begin{itemize}
	\item As before, the process {\OSPF} is the main process of the protocol. It reads messages from the message queue and calls other processes depending on message types; it periodically broadcasts \hellomsgs, checks liveness of discovered neighbours, and refreshes the router's own \lsadv.
	\item The process {\HELLO} describes all the actions performed when a \hellomsg is received. 
This includes updating the relevant inactivity timer and, if the \hellomsg is from a previously unknown neighbour, updating the node's {neighbour list}. 
The process also generates a \dbdmsg, which is sent  back to the originator of the \hellomsg, and 
a new \lsadv when required.
\item The process {\DBD} handles incoming \dbdmsgs. 
Sending requests for those \lsadvs from \dbdmsg that do not have a matching partner in the node's LSDB
is among its actions.
	\item The new process {\SNMIS} describes the actions required when a sequence number mismatch occurs during the adjacency-establishment procedure. The actions include reinitialising neighbour structures and restarting the adjacency-establishment procedure.
\item The process {\REQ} describes the actions following the receipt of an \lsrmsg, such as
 finding the requested \lsadvs in the node's local data, and sending them back to the sender of the message.
\item The process {\UPDp} manages incoming \lsumsgs, including the installation of  up-to-date \lsadvs in the node's LSDB, 
and broadcasting the updated information.	\item The process \keyw{ACK} describes the actions taken to handle the receipt of an \lsadv. This involves removing acknowledged \lsadvs from link state retransmission lists.
\item The process {\QMSG} models a message queue. As before we have two instances, a queue for incoming messages and one for 
outgoing processes.
\item The process {\QSND} is responsible for sending outgoing messages that were generated by the other processes described above.
\end{itemize}

\noindent
The system initialisation is identical to the simple model: each node runs four processes in~parallel.\\[2mm]
\centerline{$
(\xi_0, \qsndP) \parl (\zeta, {\QMSG}(\msgs)) \parl (\chi, {\OSPF}(\ip,\nbrs,\lsdb,\hellot)) \parl (\theta,{\QMSG}(\msgs))\,,
$}\\[2mm]
where the valuation functions are defined in a straightforward way; details can be found in \autoref{sec:init_simp}. 

\subsubsection{The Basic Routine}
The basic process {\OSPF} (\autoref{pro:detailed_ospf}) consists of seven parts:
handling incoming messages and distributing them to subprocesses, 
sending out \hellomsgs periodically, 
removal of neighbours that are considered inactive, 
finding neighbours whose DD timer has fired,
retransmitting {\sc Dbd}, {\sc Lsr} and  \lsumsgs,
as well as the generation of new \lsadvs.

The message handling is described in Lines \ref{detailed_simp_1} to \ref{detailed_simp_14}. 
The process first receives a message by \textbf{receive}(\msg); by our specification of the input queue\footnote{The input queue is the same for both models; see \autoref{pro:simple_message_queue}.} this will be the message currently in the front of the queue (the oldest message). 
{\OSPF} then checks the type of the message and calls the corresponding process:
in case of a \hellomsg the process {\HELLO} is called, etc. 

The second part of {\OSPF} (Lines \ref{detailed_simp_15} to \ref{detailed_simp_18}) ensures that \hellomsgs are sent periodically. 
The parameter $\hellot$ indicates when the next \hellomsg should be sent.
The process checks whether it is time to send the next \hellomsg;
 if so it resets the timer (Line \ref{detailed_simp_16}) and  broadcasts a \hellomsg. 
 In fact the message is sent to the process {\QSND} (\autoref{pro:simple_sending_queue}), which handles the actual sending. 
 
The third part of {\OSPF} (Lines \ref{detailed_simp_19} to \ref{detailed_simp_25}) collects all of the neighbours from which no activity has been seen for the last {\rtdeadintvl} seconds, removes them from {\nbrs} and informs the neighbours about these updates.
The only difference to \autoref{pro:simple_basic} is that the link state retransmission list needs to be updated (Line \ref{detailed_simp_23}), 
and that the destinations of the \lsumsg are determined by \fnfloodNIPS.
\pagebreak

  \algsetup{linenodelimiter=.,linenosize=\tiny}
  \begin{algorithm}[H]
    {
     \caption{The Basic Routine}
      \label{pro:detailed_ospf}
      \begin{algorithmic}[1]
\DEFPROCESS{\OSPF}{\ip\comma\nbrs\comma\lsdb\comma\hellot}
	\IFempty									
		\receiveL{\msg}\ .																																			\label{detailed_simp_1}
		\COMLINE{depending on the message, the node calls different processes}													\label{detailed_simp_2}
		\PAR																																								\label{detailed_simp_3}
			\IF[\hellomsg received]{$\msg = \hello{\ips}{\sip}$}																						\label{detailed_simp_4}
				\helloL\ips\sip\ip\nbrs\lsdb\hellot																											\label{detailed_simp_5}
			\ELSIF[\dbdmsg received]{$\msg = \dbdDetailed{\lsahdrs}{\sqn}{\ibit}{\sip}$}												\label{detailed_simp_6}
				\dbdDetailedL\lsahdrs\sqn\ibit\sip\ip\nbrs\lsdb\hellot																				\label{detailed_simp_7}
			\ELSIF[\lsrmsg received]{$\msg = \req{\lsahdr}{\sip}$}																				\label{detailed_simp_8}
				\reqL\lsahdr\sip\ip\nbrs\lsdb\hellot																											\label{detailed_simp_9}
			\ELSIF[\lsumsg received]{$\msg = \upd\lsas\sip$}																						\label{detailed_simp_10}
				\updL\lsas\sip\ip\nbrs\lsdb\hellot																											\label{detailed_simp_11}
			\ELSIF[\ackmsg received]{$\msg = \ack\lsahdrs\sip$}																				\label{detailed_simp_12}
				\ackL\lsahdrs\sip\ip\nbrs\lsdb\hellot																										\label{detailed_simp_13}
			\ENDIFii
		\ENDPAR																																						\label{detailed_simp_14}
	\ELSIF[send \hellomsg]{$\now\geq\hellot$}																										\label{detailed_simp_15}
		\UPD{\hellot := \now + \hellointvl}																													\label{detailed_simp_16}
		\sendL{\sndmsg{\hello{\{\pi_1(n) \mid n \in\nbrs\}}{\ip}}{\emptyset}}\ .																\label{detailed_simp_17}
		\ospfL\ip\nbrs\lsdb\hellot																																\label{detailed_simp_18}
	\ELSIF[inactive neighbours, which should be removed, exist]{$\deadNBRS{\nbrs}{\now} \neq \emptyset$}			\label{detailed_simp_19}
		\UPD{\nbrs := \nbrs - \deadNBRS\nbrs\now}	\COM{remove corresponding neighbour structures}					\label{detailed_simp_20}
		\UPD{\lsa := \newLSA\ip\now\nbrs}	\COM{generate a new \lsadv}																\label{detailed_simp_21}
		\UPD{\lsdb := \install\lsdb{\{\lsa\}}}		\COM{install it}																						\label{detailed_simp_22}
		\UPD{\nbrs := \updRXMTS\nbrs{\{\lsa\}}	}	\COM{update the link state retransmission list}								\label{detailed_simp_23}
		\sendL{\sndmsg{\upd{\{\lsa\}}{\ip}}{\floodNIPS{\nbrs}}}\ .\COM{and flood it out}												\label{detailed_simp_24}
		\ospfL\ip\nbrs\lsdb\hellot																																\label{detailed_simp_25}		
	\ELSIF[find neighbour whose DD timer has fired]{$\nip = \ddNIP\nbrs\now\ip$}													\label{detailed_simp_26}	
		\UPD{\nbrs := \setDDT\nbrs\nip{\now{+}\rxmtintvl}}																						\label{detailed_simp_27}	
		\sendL{\sndmsg{\genDBD\nbrs\lsdb\nip\ip}{\{\nip\}}}\ . 																					\label{detailed_simp_28}	
		\ospfL\ip\nbrs\lsdb\hellot																																\label{detailed_simp_29}	
	\ELSIF[find a neighbour whose request timer has fired]{$\nip = \reqNIP\nbrs\now$}											\label{detailed_simp_30}	
		\UPD{\nbrs := \setREQT{\nbrs}{\nip}{\now{+}\rxmtintvl}}																					\label{detailed_simp_31}	
		\sendL{\sndmsg{{\req{\sel{\getREQS{\nbrs}{\nip}}}{\ip}}}{\{\nip\}}}\ . 																\label{detailed_simp_32}	
		\ospfL\ip\nbrs\lsdb\hellot																																\label{detailed_simp_33}	
	\ELSIF[find a neighbour whose retransmission timer has fired]{$\nip = \rxmtNIP\nbrs\now$}								\label{detailed_simp_34}	
		\UPD{\nbrs := \setRXMTS{\nbrs}{\nip}{\now{+}\rxmtintvl}}																					\label{detailed_simp_35}	
		\sendL{\sndmsg{\upd{\getRXMTS\nbrs\nip}{\ip}}{\{\nip\}}}\ .																			\label{detailed_simp_36}	
		\ospfL\ip\nbrs\lsdb\hellot																																\label{detailed_simp_37}	
	\ELSIF[self-originated \lsadv too old]{$\neg\lsaExist{\lsdb}{(\ip, \now - \refreshintvl)}$}										\label{detailed_simp_38}	
		\UPD{\lsa := \newLSA\ip\now\nbrs}\COM{generate new one}																			\label{detailed_simp_39}	
		\UPD{\lsdb := \install\lsdb{\{\lsa\}}}																													\label{detailed_simp_40}	
		\UPD{\nbrs := \updRXMTS{\nbrs}{\{\lsa\}}}																										\label{detailed_simp_41}	
		\sendL{\sndmsg{\upd{\{\lsa\}}{\ip}}{\floodNIPS\nbrs}}\ . 																					\label{detailed_simp_42}	
		\ospfL\ip\nbrs\lsdb\hellot																																\label{detailed_simp_43}	
\ENDIFii

		\end{algorithmic}
    }
  \end{algorithm}

Lines \ref{detailed_simp_26} to \ref{detailed_simp_29} handle the retransmission of \dbdmsgs every {\rxmtintvl} seconds.
The function $\ddNIP\nbrs\now\ip$ returns the IP address of a neighbour whose DD timer has fired.
The DD timer for that neighbour is reset (Line \ref{detailed_simp_27}) and a \dbdmsg is sent to it.

Similarly, Lines \ref{detailed_simp_30} to \ref{detailed_simp_33} handle the retransmission of \lsrmsgs every \textit{rxmtintvl} seconds.
Instead of a \dbdmsg, a \lsrmsg is sent out requesting a new \lsadv.

In the same spirit, Lines \ref{detailed_simp_37} to \ref{detailed_simp_36} handle the retransmission of \lsumsgs every \textit{rxmtintvl} seconds.

The last part of {\OSPF}, Lines \ref{detailed_simp_38} to \ref{detailed_simp_43}, is executed if there is an \lsadv in the link state database that is too old.
In that case the process generates  a new \lsadv in Line \ref{detailed_simp_39} (which happens every \textit{refreshintvl} seconds),
installs that \lsadv into its {\lsdb} (Line \ref{detailed_simp_40}), updates the neighbour list accordingly (Line \ref{detailed_simp_41}), and floods the \lsadv out (Line \ref{detailed_simp_42}).

\subsubsection{Message Processing}

The process {\HELLO} (\autoref{pro:detailed_hello_app}) handles the receipt of a \hellomsg. 

  \algsetup{linenodelimiter=.,linenosize=\tiny}
  \begin{algorithm}[H]
    {
     \caption{Handling \hellomsgs}
      \label{pro:detailed_hello_app}
      \begin{algorithmic}[1]
\DEFPROCESS{\HELLO}{\ips\comma\sip\comma\ip\comma\nbrs\comma\lsdb\comma\hellot}
	\IF[the sender {\sip} is unknown]{$\neg\nbrExist\nbrs\sip$}																\label{hello_det_app_1}
		\UPD{\nbrs := \newNBR{\nbrs}{\sip}}																							\label{hello_det_app_2}
		\helloL\ips\sip\ip\nbrs\lsdb\hellot																								\label{hello_det_app_3}
	\ELSIF[{\sip} is a known neighbour]{$\nbrExist{\nbrs}{\sip}$}															\label{hello_det_app_4}
		\UPD{\nbrs := \setINACTT{\nbrs}{\sip}{\now+\rtdeadintvl}}															\label{hello_det_app_5}
		\UPD{\ns := \getNS{\nbrs}{\sip}}																									\label{hello_det_app_6}
		\PAR																																			
		\IF[2-WayReceived, start adjacency-forming]{$\ip\in\ips \land \adj\ip\sip \land \ns = \str{Init}$}		\label{hello_det_app_8}
			\UPD{\nbrs := \setNS{\nbrs}{\sip}{\str{ExStart}}}																		\label{hello_det_app_9}
			\UPD{\nbrs := \incDDSQN\nbrs\sip}																						\label{hello_det_app_10}
			\UPD{\nbrs := \setDDT{\nbrs}{\sip}{\now + \rxmtintvl}}															\label{hello_det_app_11}
			\sendL {\sndmsg{\genDBD{\nbrs}{\lsdb}{\sip}{\ip}}{\{\sip\}}}\ .\ 			\label{hello_det_app_12}
			\ospfL{\ip}{\nbrs}{\lsdb}{\hellot}																						\label{hello_det_app_13}
		\ELSIF[Adjacency-forming already underway]{$\ip\in\ips \land \adj\ip\sip \land \ns\geq\str{ExStart}$}							\label{hello_det_app_14}
			\ospfL\ip\nbrs\lsdb\hellot																										\label{hello_det_app_15}
		\ELSIF[2-WayReceived, adjacency-forming not needed]{$\ip\in\ips \land \neg\adj\ip\sip$}			\label{hello_det_app_16}
			\UPD{\nbrs := \setNS{\nbrs}{\sip}{\str{2-Way}}}																		\label{hello_det_app_17}
			\ospfL{\ip}{\nbrs}{\lsdb}{\hellot}																								\label{hello_det_app_18}
		\ELSIF[1-WayReceived]{$\ip\notin\ips$}																						\label{hello_det_app_19}
			\UPD{\nbrs := \initNBR{\nbrs}{\sip}{\str{Init}}}																			\label{hello_det_app_20}
			\ospfL{\ip}{\nbrs}{\lsdb}{\hellot}																								\label{hello_det_app_21}
		\ENDIFii	
		\ENDPAR																																	\label{hello_det_app_22}
	\ENDIFii

		\end{algorithmic}
    }
  \end{algorithm}

\vspace{-1pt}

The first part of the process (Lines \ref{hello_det_app_1} to \ref{hello_det_app_3}) is executed
when the originator of the \hellomsg is unknown to the node {\ip}. The neighbour list is updated in Line~\ref{hello_det_app_2} 
and the process calls itself. As the node is now aware of its neighbour \sip, Lines \ref{hello_det_app_4} to \ref{hello_det_app_22} 
are executed.

In the second part, Lines \ref{hello_det_app_4} ff., the sender {\sip} of the \hellomsg is already known.
As the node communicates with $\sip$ the corresponding inactivity timer is reset in Line \ref{hello_det_app_5}. 
The subsequent actions depend on the state of the node.
If the sender has discovered the node $\ip$ ($\ip\in\ips$) and the two nodes are intended to form an adjacency 
(\adj\sip\ip) but the establishment process has not yet started (the status is 
\str{Init}) then the node initialises the exchange of the LSDBs. For that it sets the neighbour status to \str{ExStart} (Line \ref{hello_det_app_9}), increments the DD sequence number (Line \ref{hello_det_app_10}) to identify the outgoing \dbdmsg (Line \ref{hello_det_app_12}) in a unique way. It also resets the DD timer in Line \ref{hello_det_app_11}.
If the adjacency-establishment procedure has already begun ($\ns\geq\str{ExStart}$)---of course the node $\sip$ should still be aware of $\ip$---no actions are required as the other processes, such as {\DBD}, handle the necessary actions (Lines \ref{hello_det_app_14} and \ref{hello_det_app_15}).
In case the node was discovered by \sip, but adjacency is prohibited ($\neg\adj\ip\sip$) then the neighbour status is changed to \str{2-Way} (Line \ref{hello_det_app_17}), indicating that a bidirectional link has been established; no other actions are necessary.
If $\ip$ has not yet been discovered by \sip, or if the connectivity between the two nodes broke down in the past, 
then the corresponding neighbour structure will be reset (or initialised) in Line \ref{hello_det_app_20}.

The process {\DBD} (\autoref{pro:detailed_dbd}) handles the receipt of a \dbdmsg. 
It is the process that differs the most compared to the simple model for it handles most of the adjacency-establishment procedure.
To increase readability we use predicates in the guards---we describe these predicates along with a detailed explanation of the steps of the process.

Line \ref{dbd_det_2} ignores the \dbdmsg in case the sender $\sip$ is unknown, which is checked in Line  \ref{dbd_det_1}.

Lines  \ref{dbd_det_3}-- \ref{dbd_det_49} handle the case where the sender $\sip$ is known. 
Depending on the status of the adjacency-establishment procedure the process performs different actions.
The decision of which actions are taken depend on six conditions:
(a) the intended adjacency relationship between $\ip$ and $\sip$ ($\adj\ip\sip$);
(b) $\ip$'s neighbour state for $\sip$;
(c) the DD timer for $\sip$;
(d) the DD sequence number;
(d) the IP addresses of {\ip} and {\sip}; and
(f) the bit {\ibit} in the received \dbdmsg.
Lines \ref{dbd_det_4} and \ref{dbd_det_5} describe the case where the nodes do not form an adjacency, and 
the nodes communicated before ($\ns=\str{Init}$). By the incoming \dbdmsg bidirectional communication is 
established; this fact is registered in the neighbour structure by setting the neighbour status for $\sip$ to 
\str{2-Way}. 
In case the status is already set to \str{2-Way} (Line \ref{dbd_det_6}), indicating that full communication has been established and 
also that adjacency is not sought, the message can be ignored (Line  \ref{dbd_det_7}).
The remaining cases concern the establishment of adjacency. 
If the nodes aim for adjacency, but the procedure has not been started ($\ns = \str{Init} \wedge \adj\ip\sip$), checked in \ref{dbd_det_7},
then the adjacency establishment begins by sending out a \dbdmsg in Line \ref{dbd_det_11} and the process {\DBD} is called again.
 Before that the node's data structures need updates: the neighbour state for {\sip} is set to \str{ExStart} (Line \ref{dbd_det_8}) showing that the adjacency procedure has indeed begun, the DD  sequence number is incremented and the DD timer reset (Lines \ref{dbd_det_8} and \ref{dbd_det_9}).
Calling the DBD process again allows a slave to immediately send another DBD message (executing Lines~\ref{dbd_det_13}  to \ref{dbd_det_15}). Determining the slave/master relationship is the first proper step in the adjacency-establishment procedure.
Although the RFC allows other methods as well, we use the total order on IP addresses as underlying criterion. 
We define abbreviations to check the slave/master status of a node $\ip$ with regards to $\sip$:\\[2mm]
\centerline{
$\keyw{is\_slave} := \adj\ip\sip \wedge \ip < \sip
\qquad\qquad
\keyw{is\_master} := \adj\ip\sip \wedge \ip > \sip
$}\\[2mm]
\noindent As we assume IP addresses to be unique for each node in a network, we can ignore the case $\ip=\sip$.

In case an adjacency establishment has begun ($\ns=\str{ExStart}$), the node itself identified itself as slave, \ie it has a smaller sequence number
than the sender, and it receives the first message from the
 master during the establishment (the bit {\ibit} is set),\footnote{%
This phase is called ``NegotiationDone" in the RFC.} then the node 
updates its neighbour structure, indicating that an \str{Exchange} of messages is happening with $\sip$, 
and the stored sequence number of that node (both \noindent Line  \ref{dbd_det_14}).
It \hfill does \hfill not \hfill reset \hfill the \hfill DD \hfill timer \hfill as \hfill  it \hfill is \hfill  up \hfill  to \hfill  the \hfill  master \hfill  to \hfill  investigate \hfill  in \hfill case

  \algsetup{linenodelimiter=.,linenosize=\tiny}
  \begin{algorithm}[H]
    {
     \caption{Handling \dbdmsgs}
      \label{pro:detailed_dbd}
      \begin{algorithmic}[1]
\small
\vspace{-2pt}
\DEFPROCESS{\DBD}{\lsahdrs\comma\sqn\comma\ibit\comma\sip\comma\ip\comma\nbrs\comma\lsdb\comma\hellot}
	\IF[the sender {\sip} is unknown]{$\neg\nbrExist\nbrs\sip$}	 																										\label{dbd_det_1}
		\ospfL\ip\nbrs\lsdb\hellot																																						\label{dbd_det_2}
	\ELSIF[\!{\sip} is known\!]{\!$\getNBR\nbrs\sip\mathop= (\sip, \ns, \inactt, \ddsqn, \ddt, \reqs, \reqt, \rxmts, \rxmtt)\!$}					\label{dbd_det_3}
		\PARIF[no need for adjacency establishment]{$\ns=\str{Init} \wedge \neg\adj\ip\sip$}																\label{dbd_det_4}
				\UPD{\nbrs := \setNS{\nbrs}{\sip}{\str{2-Way}}}\ \ospfP\ip\nbrs\lsdb\hellot																		\label{dbd_det_5}
			\ELSIF[ignore the message]{$\ns = \str{2-Way}$}																													\label{dbd_det_6}
				\ospfL\ip\nbrs\lsdb\hellot																																				\label{dbd_det_7}
			\ELSIF[start adjacency-forming]{$\ns = \str{Init} \wedge \adj\ip\sip$}																					\label{dbd_det_8}
				\UPD{\nbrs := \setNS{\nbrs}{\sip}{\str{ExStart}}}
				\UPDP{\nbrs := \incDDSQN{\nbrs}{\sip}}																														\label{dbd_det_9}
				\UPD{\nbrs := \setDDT{\nbrs}{\sip}{\now{+}\rxmtintvl}}																									\label{dbd_det_10}
				\sendL{\sndmsg{\genDBD{\nbrs}{\lsdb}{\sip}{\ip}}{\{\sip\}}}\ . 																							\label{dbd_det_11}
				\dbdDetailedL\lsahdrs\sqn\ibit\sip\ip\nbrs\lsdb\hellot																										\label{dbd_det_12}
			\ELSIF[negotiation done]{$\keyw{negotiate\_slave}$}																											\label{dbd_det_13}
				\UPD{\nbrs := \setNS\nbrs\sip{\str{Exchange}}}
				\UPDP{\nbrs := \setDDSQN\nbrs\sip\sqn}																														\label{dbd_det_14}
				\sendL{\sndmsg{\genDBD{\nbrs}{\lsdb}{\sip}{\ip}}{\{\sip\}}}\ .\ \ospfP\ip\nbrs\lsdb\hellot													\label{dbd_det_15}
			\ELSIF[negotiation done]{$\keyw{negotiate\_master}$}																										\label{dbd_det_16}
				\UPD{\nbrs := \setNS\nbrs\sip{\str{Exchange}}}
				\UPDP{\nbrs := \incDDSQN\nbrs\sip}																																\label{dbd_det_17}
				\UPD{\nbrs := \setDDT\nbrs\sip{\now{+}\rxmtintvl}}
				\UPDP{\nbrs := \addREQS\nbrs\sip\lsdb\lsahdrs}																											\label{dbd_det_18}
				\sendL{\sndmsg{\genDBD\nbrs\lsdb\sip\ip}{\{\sip\}}}\ . \ \ospfP\ip\nbrs\lsdb\hellot															\label{dbd_det_19}
			\ELSIF[ignore the message]{$\keyw{negotiate\_others}$}																										\label{dbd_det_20}
				\ospfL\ip\nbrs\lsdb\hellot																																				\label{dbd_det_21}
			\ELSIF[duplicate message]{$\keyw{exchange\_duplicate\_slave}$}																						\label{dbd_det_22}
				\sendL{\sndmsg{\genDBD\nbrs\lsdb\sip\ip}{\{\sip\}}}\ .\ \ospfP\ip\nbrs\lsdb\hellot																\label{dbd_det_23}
			\ELSIF[duplicate message]{$\keyw{exchange\_duplicate\_master}$}																						\label{dbd_det_24}
				\ospfL\ip\nbrs\lsdb\hellot																																				\label{dbd_det_25}
			\ELSIF[exchange done]{$\keyw{exchange\_slave}$}																											\label{dbd_det_26}
				\UPD{\nbrs := \incDDSQN\nbrs\sip}
				\UPDP{\nbrs := \setDDT\nbrs\sip{\now{+}\rtdeadintvl}}																									\label{dbd_det_27}
				\UPD{\nbrs := \addREQS\nbrs\sip\lsdb\lsahdrs}																												\label{dbd_det_28}
				\sendL{\sndmsg{\genDBD\nbrs\lsdb\sip\ip}{\{\sip\}}}\ .																									\label{dbd_det_29}
				\PARIF{$\getREQS\nbrs\sip \neq\emptyset$}																										\label{dbd_det_30}
					\UPD{\nbrs := \setNS{\nbrs}{\sip}{\str{Loading}}}\ .\ \ospfP\ip\nbrs\lsdb\hellot																\label{dbd_det_31}
				\ELSIF{$\getREQS\nbrs\sip = \emptyset$}																											\label{dbd_det_32}
					\UPD{\nbrs := \setNS\nbrs\sip{\str{Full}}}
					\UPDP{\lsa := \newLSA\ip\now\nbrs}																															\label{dbd_det_33}
					\UPD{\lsdb := \install\lsdb{\{\lsa\}}}
					\UPDP{\nbrs := \updRXMTS\nbrs{\{\lsa\}}}																													\label{dbd_det_34}
					\sendL{\sndmsg{\upd{\{\lsa\}}{\ip}}{\floodNIPS\nbrs}}\ .\ \ospfP\ip\nbrs\lsdb\hellot														\label{dbd_det_35}
				\ENDPARIF
			\ELSIF[exchange done]{$\keyw{exchange\_master}$}																										\label{dbd_det_36}
				\UPD{\nbrs := \incDDSQN\nbrs\sip}
				\UPDP{\nbrs := \addREQS\nbrs\sip\lsdb\lsahdrs}																											\label{dbd_det_37}
				\PARIF{$\getREQS\nbrs\sip\neq\emptyset$}																											\label{dbd_det_38}
						\UPD{\nbrs := \setNS\nbrs\sip{\str{Loading}}}\ .\ \ospfP\ip\nbrs\lsdb\hellot																\label{dbd_det_39}
					\ELSIF{$\getREQS\nbrs\sip=\emptyset$}																											\label{dbd_det_40}
						\UPD{\nbrs := \setNS\nbrs\sip{\str{Full}}}
						\UPDP{\lsa := \newLSA\ip\now\nbrs}																														\label{dbd_det_41}
						\UPD{\lsdb := \install\lsdb{\{\lsa\}}}
						\UPDP{\nbrs := \updRXMTS\nbrs{\{\lsa\}}}																												\label{dbd_det_42}
						\sendL{\sndmsg{\upd{\{\lsa\}}{\ip}}{\floodNIPS\nbrs}}\ .\ \ospfP\ip\nbrs\lsdb\hellot													\label{dbd_det_43}
					\ENDPARIF
			\ELSIF[sqn mismatch sqn]{$\keyw{exchange\_others}\vee \keyw{load\_others}$} 												\label{dbd_det_44}
				\snmisL\sip\ip\nbrs\lsdb\hellot																																		\label{dbd_det_45}
			\ELSIF[duplicate message]{$\keyw{load\_duplicate\_slave }$}																								\label{dbd_det_46}
				\sendL{\sndmsg{\genDBD{\nbrs}{\lsdb}{\sip}{\ip}}{\{\sip\}}}\ .\ \ospfP\ip\nbrs\lsdb\hellot														\label{dbd_det_47}
			\ELSIF[duplicate message]{$\keyw{load\_duplicate\_master}$}																								\label{dbd_det_48}
				\ospfL\ip\nbrs\lsdb\hellot																																				\label{dbd_det_49}
			\ENDPARIF
\ENDIFii

		\end{algorithmic}
    }
  \end{algorithm}

\noindent some \dbdmsgs are not acknowledged.
Eventually, in Line  \ref{dbd_det_15}, the node $\ip$ replies with another \dbdmsg, which contains the information of its own \lsdb.
The check of \ref{dbd_det_13}, described above, is formally defined as
\vspace{-2mm}
\[
	\keyw{negotiate\_slave} := (\ns = \str{ExStart}) \wedge \keyw{is\_slave} \wedge \ibit\,.
\]
Lines  \ref{dbd_det_16}-- \ref{dbd_det_19} is the counterpart of Lines  \ref{dbd_det_13}-- \ref{dbd_det_15} in case 
the node identifies itself as master. The check in Line \ref{dbd_det_16} is defined by\\[1mm]
\centerline{$
\keyw{negotiate\_master} := (\ns = \str{ExStart}) \wedge \keyw{is\_master} \wedge (\sqn = \ddsqn) \wedge \neg\ibit\,.
$}\\[2mm]
The node does not only check whether the adjacency establishment has started and whether it is the master in the procedure, 
it also checks that the sequence number of the incoming message matches its expectation and that its bit is set to false.
These two conditions intuitively mean that the slave acknowledges the current node as master.
In case the guard of Line \ref{dbd_det_16} evaluates to true, the neighbour state for {\sip} is set to \str{Exchange}
and the DD sequence number is incremented (Line \ref{dbd_det_17}).
Moreover, the DD timer is reset and any \lsadv header listed in the incoming message that is more recent than the corresponding one in the node's own LSDB are added to the link state request list (both Line  \ref{dbd_det_18}).
Lastly, the node sends a \dbdmsg back to the sender {\sip} (Line  \ref{dbd_det_19}). 
All other cases where the neighbour state is \str{Init} are ignored (\ref{dbd_det_21}), the guard of Line  \ref{dbd_det_20}) is
\vspace{-2mm}
\[
	\keyw{negotiate\_others} := (\ns = \str{ExStart}) \wedge \neg\keyw{negotiate\_slave} \wedge \neg\keyw{negotiate \_master}\,.
\]
Lines \ref{dbd_det_22} --\ref{dbd_det_25} handle situations where a \dbdmsg is received a second time, which can happen due to loss of 
messages acknowledging other messages.
If the slave receives a \dbdmsg a second time ($\sqn \leq \ddsqn$) from the master, it assumes that its reply, sent in Line \ref{dbd_det_15}, was lost and 
resends the message (Line \ref{dbd_det_23}).
Formally the guard is \vspace{-2mm}
\[
	\keyw{exchange\_duplicate\_slave} :=  (\ns = \str{Exchange}) \wedge  \keyw{is\_slave} \wedge (\sqn \leq \ddsqn)\,.
\]
In case the master receives a duplicate of a message, checked in Line \ref{dbd_det_24} using  \vspace{-2mm}
\[
	\keyw{exchange\_duplicate\_master} :=  (\ns = \str{Exchange}) \wedge \keyw{is\_master} \wedge (\sqn < \ddsqn)\,,
\]
no actions are required and the message is disregarded.
In Line  \ref{dbd_det_26} the node, which partakes in a adjacency-establishment procedure as slave, 
handles a message from the master as answer to the \dbdmsg sent earlier.
\[ 
	\keyw{exchange\_slave} :=  (\ns = \str{Exchange}) \wedge \keyw{is\_slave} \wedge (\sqn = \ddsqn{+}1) \wedge \neg\ibit
\]
It enters the next phase of the adjacency-forming procedure, called ``ExchangeDone" in the RFC.
In this phase the slave receives a \dbdmsg presenting the LSDB of the master. 
As a consequence, DD sequence number is incremented and the DD timer is reset (Line~\ref{dbd_det_27}).
Moreover,  any {\lsadv} header that is more recent than the corresponding on in the node's own database is added to the link state request list 
(Line \ref{dbd_det_28}). After the node updated its local data, it sends another \dbdmsg back to the master {\sip} (Line \ref{dbd_det_29}).

After that, the node checks its request list, \ie it checks whether there are some \lsadvs in the master's LSDB that are needed by $\ip$.
In case there are such \lsadvs (Line \ref{dbd_det_30})
the node enters the next stage of the adjacency establishment by setting  the neighbour state to \str{Loading}, 
and returns to main process {\OSPF}, waiting for $\sip$ to send the necessary data.
In case there are no \lsadvs in the master's LSDB that are needed by the node itself (Line \ref{dbd_det_32}), 
the slave $\sip$ considers the establishment procedure finished and sets the status to \str{Full} (Line \ref{dbd_det_33}).
It also updates its own data structures by 
generating a new \lsadv (Line \ref{dbd_det_32}), installing it in its own \lsdb\ (Line \ref{dbd_det_33}) and flooding it out (Line \ref{dbd_det_35}).

If the master receives a \dbdmsg from the slave containing information it is waiting for, \ie
\[\keyw{exchange\_master} :=  (\ns = \str{Exchange}) \wedge \keyw{is\_master} \wedge (\sqn = \ddsqn) \wedge \neg\ibit\,,\]
it enters the next phase of the adjacency-forming procedure, which again is called ``ExchangeDone" in the RFC.
In that case, the master increments the corresponding sequence number and updates its link state request list, as usual (Line \ref{dbd_det_37}).
Similar to the previous case the node checks whether there are some \lsadvs in the slave's LSDB that the master requires.
If there are \lsadvs missing, \ie the request list is not empty (Line \ref{dbd_det_38}), the neighbour state is changed to \str{Loading} (Line \ref{dbd_det_39}), indicating that the slave will send the missing information in the future.
In case there are no \lsadvs missing (Line \ref{dbd_det_40}), the master considers the adjacency-establishment procedure to be done, 
indicates this by setting the neighbour status to \str{Full} (Line \ref{dbd_det_41}) updates its data structures in the same way as the slave did (see above),
and, identical to the slave, informs its neighbours about the new available information, by sending out an {\lsadv} (Line \ref{dbd_det_43}).
Afterwards, the node returns to the main process {\OSPF}.

In case that during the establishment procedure either the slave or the master receives a message out of order, which means that the 
sequence number in the \dbdmsg does not fit expectations, the process {\SNMIS} is called, which handles this mismatch.
To check whether the incoming message is out of order we use, in Line \ref{dbd_det_44},
\vspace{-2mm}
\[\begin{array}{l@{\ }l}
	\keyw{exchange\_others} := &(\ns = \str{Exchange}) \wedge \neg\keyw{exchange\_slave} \wedge \neg\keyw{exchange\_master}\\
			& \wedge\ \neg\keyw{exchange\_duplicate\_master} \wedge \neg\keyw{exchange\_duplicate\_slave}\,.
			\end{array}\]

Lines \ref{dbd_det_46}--\ref{dbd_det_49} handle the cases where a slave (master) receives a message it handled before
while being in phase \str{Loading} or \str{Full} of the adjacency-establishment procedure.
For the slave and the master this is checked by 
\[\begin{array}{@{}r@{\ }c@{\ }l@{}}
	\keyw{load\_duplicate\_slave} &:=  & (\ns \geq \str{Loading}) \wedge \keyw{is\_slave} \wedge (\sqn \leq \ddsqn) \wedge \!\neg\ibit \wedge \ddt \geq \now\ \text{ and}\\
	\keyw{load\_duplicate\_master} &:= & (\ns \geq \str{Loading}) \wedge \keyw{is\_master} \wedge (\sqn < \ddsqn) \wedge \neg\ibit\,,
\end{array}\]
respectively. 
If the slave receives a duplicate (Line \ref{dbd_det_46}), it responds with the same \dbdmsg it had sent before in Line \ref{dbd_det_29}, 
assuming that its message was lost; this happens in Line \ref{dbd_det_47}.
  In case the master handles a duplicate (Line \ref{dbd_det_48}) then no actions are required, the node returns to the main process 
  {\OSPF} (Line \ref{dbd_det_49}).
  
 The only remaining case is the neighbour status is \str{Loading} or \str{Full}, but the message is neither expected nor a duplicate. 
 In short that means that something went wrong during the adjacency-establishment procedure.
 We check this situation using the guard\\[2mm]
 \centerline{$
 	\keyw{load\_others} :=  (\ns \geq \str{Loading}) \wedge \neg\keyw{load\_duplicate\_slave} \wedge \neg\keyw{load\_duplicate\_slave}\,.
	$}\\[2mm]
As in the case of an out-of-order message during the message-exchange phase (Lines \ref{dbd_det_44} and \ref{dbd_det_45}), 
the process calls {\SNMIS} to handle this situation. As these two scenarios are handled 
in the same way, we use the combined guard $\keyw{load\_others} \vee \keyw{exchange\_others}$ in Line \ref{dbd_det_44}.

\begin{figure}[t]
	\centering
	\includegraphics[width=.6\textwidth]{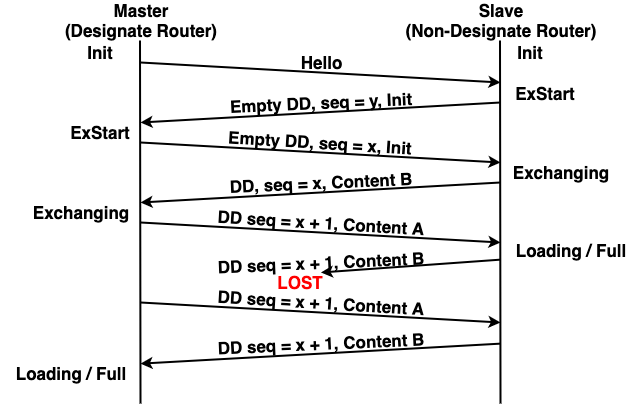}
	\caption{Communication between master and slave during adjacency establishment\label{fig:ex_pingpong}}
	\vspace{-3mm}
\end{figure}

\autoref{fig:ex_pingpong} summarises the (somewhat complicated) situation of sending messages forth and back between master and slave.
It illustrates a possible scenario in which \dbdmsgs are exchanged by two routers.
Time increases from the top to the bottom, each arrow indicates a message sent, the words above each arrow represent the message type and contents and the words on the left and the right of the figure represent  the node's neighbour status of the other node.

The process \keyw{SNMIS} (\autoref{pro:detailed_snmis}) contains all actions that are taken when something goes wrong during the adjacency-forming procedure for a node and its neighbouring router \sip. 
Such an error, called ``SeqNumberMismatch" in the RFC, effectively restarts the adjacency-establishment procedure. 

  \algsetup{linenodelimiter=.,linenosize=\tiny}
  \begin{algorithm}[H]
    {
     \caption{Sequence Numbers Mismatches}
      \label{pro:detailed_snmis}
      \begin{algorithmic}[1]
\DEFPROCESS{SNMIS}{\sip\comma\ip\comma\nbrs\comma\lsdb\comma\hellot}
	\UPD{\nbrs := \initNBR\nbrs\sip{\str{ExStart}}}										\label{snmis_det_1}
	\UPD{\nbrs := \incDDSQN\nbrs\sip}														\label{snmis_det_2}
	\UPD{\nbrs := \setDDT\nbrs\sip{\now{+}\rxmtintvl}}									\label{snmis_det_3}
	\sendL{\sndmsg{\genDBD\nbrs\lsdb\sip\ip}{\{\sip\}}}\ . 							\label{snmis_det_4}
	\ospfL\ip\nbrs\lsdb\hellot																		\label{snmis_det_5}
		\end{algorithmic}
    }
  \end{algorithm}

\noindent The process  resets the status of the adjacency-establishment to \str{ExStart} (Line  \ref{snmis_det_1}), 
increments the DD sequence number to indicate a newer forthcoming message (Line \ref{snmis_det_2}) and, 
as usual, resets the DD timer (Line \ref{snmis_det_1}). 
It then restarts the establishment procedure by sending out a new \dbdmsg, similar to Line~\ref{dbd_det_11} of \autoref{pro:detailed_dbd}.

The process \keyw{REQ} (\autoref{pro:detailed_req}) involves all actions that are taken upon the receipt of a \lsrmsg.

  \algsetup{linenodelimiter=.,linenosize=\tiny}
  \begin{algorithm}[H]
    {
     \caption{Handling \lsrmsgs}
      \label{pro:detailed_req}
      \begin{algorithmic}[1]
\DEFPROCESS{\REQ}{\lsahdr\comma\sip\comma\ip\comma\nbrs\comma\lsdb\comma\hellot}
	\COMLINE{the sender {\sip} is ``unknown''}
	\IF{$\neg\nbrExist\nbrs\sip \vee \getNS\nbrs\sip < \str{Exchange} \vee \neg\lsaExist\lsdb\lsahdr$}		\label{req_det_2}
		\ospfL\ip\nbrs\lsdb\hellot 																												\label{req_det_3}
	\STATE \hspace*{-3em}\COM{the sender {\sip} is ``known''}																\label{req_det_4}
	\ELSIF{$\nbrExist\nbrs\sip \wedge \getNS\nbrs\sip\geq\str{Exchange} \wedge\lsaExist\lsdb\lsahdr$}	\label{req_det_5}
		\sendL{\sndmsg{\upd{\{\getLSA\lsdb\lsahdr\}}{\ip}}{{\{\sip\}}}}\ . 
		\ospfP\ip\nbrs\lsdb\hellot 																												\label{req_det_6}
	\ENDIFii
		\end{algorithmic}
    }
  \end{algorithm}

\noindent It is nearly identical to the simplified version, depicted in \autoref{pro:simple_req}, only the guards are more complex as 
they have to reflect the status of the adjacency-establishment procedure.
Line~\ref{req_det_2} checks whether the sender of {\sip} of the \lsrmsg is unknown, is in a premature state ($\ns<\str{Exchange}$) where is should not send a request, or whether more recent information was already received. In all these cases the message is dropped (Line~\ref{req_det_3}).
Line~\ref{req_det_5} is the complimentary guard: the node handles the request, looks up information in its own link state database and 
replies with an \lsumsg to \sip.

The process {\UPDp} (\autoref{pro:detailed_upd}) includes all actions that should be taken upon receipt of a \lsumsg.

  \algsetup{linenodelimiter=.,linenosize=\tiny}
  \begin{algorithm}[H]
    {
     \caption{Handling \lsumsgs}
      \label{pro:detailed_upd}
      \begin{algorithmic}[1]
\DEFPROCESS{\UPDp}{\lsas\comma\sip\comma\ip\comma\nbrs\comma\lsdb\comma\hellot}
	\IF[sender {\sip} unknown] {$\neg\nbrExist\nbrs\sip$}																									\label{upd_det_1}
		\ospfL\ip\nbrs\lsdb\hellot 																																		\label{upd_det_2}
	\ELSIF[sender {\sip} known]{$\nbrExist\nbrs\sip$}																										\label{upd_det_3}
		\sendL{\sndmsg{\ack{\{\hdr{lsa} \mid lsa\in \lsas\}}{\ip}}{\{\sip\}}}\ .																			\label{upd_det_4}
		\UPD{\lsas := \{lsa \mid lsa \in \lsas \wedge \nexists\, lsa' \in\lsdb : \hdr{lsa}\leq \hdr{lsa'}\}}\COM{\lsadvs to update LSDB}		\label{upd_det_5}
		\PAR																																										\label{upd_det_6}
			\IF{$\lsas = \emptyset$}																																		\label{upd_det_7}
				\ospfL\ip\nbrs\lsdb\hellot 																																\label{upd_det_8}
			\ELSIF{$\lsas \neq \emptyset$}																															\label{upd_det_9}
				\UPD{\lsdb := \install\lsdb\lsas}																														\label{upd_det_10}
				\UPD{\nbrs := \cleanREQS\nbrs\sip\lsdb}																										\label{upd_det_11}
				\UPD{\nbrs := \updRXMTS\nbrs\lsas}																											\label{upd_det_12}
				\sendL{\sndmsg{\upd{\lsas}{\ip}}{\floodNIPS{\nbrs}}}\ .																					\label{upd_det_13}
				\PAR																																								\label{upd_det_14}
					\IF{$\getREQS\nbrs\sip\neq \emptyset$}																							\label{upd_det_15}
						\ospfL\ip\nbrs\lsdb\hellot 																														\label{upd_det_16}
					\ELSIF{$\getREQS\nbrs\sip = \emptyset$}																						\label{upd_det_17}
						\UPD{\nbrs := \setNS\nbrs\sip{\str{Full}}}																								\label{upd_det_18}
						\UPD{\lsa := \newLSA\ip\now\nbrs}																										\label{upd_det_19}
						\UPD{\lsdb := \install\lsdb{\{\lsa\}}}																											\label{upd_det_20}
						\UPD{\nbrs := \updRXMTS\nbrs{\{\lsa\}}}																								\label{upd_det_21}
						\sendL{\sndmsg{\upd{\{\lsa\}}{\ip}}{\floodNIPS\nbrs}}\ . 																			\label{upd_det_22}
						\ospfL\ip\nbrs\lsdb\hellot 																														\label{upd_det_23}
					\ENDIFii
				\ENDPAR																																						\label{upd_det_24}
			\ENDIFii
		\ENDPAR																																								\label{upd_det_25}
	\ENDIFii
		\end{algorithmic}
    }
  \end{algorithm}

\vspace{-2mm}

As in many other processes we first handle the case that the sender of the message is unknown (Line~\ref{upd_det_1}); 
as usual the message is ignored.

Lines \ref{upd_det_3} to \ref{upd_det_25} handle the situation where {\sip} is known.
First, in Line \ref{upd_det_4} the node replies to {\sip} by sending an \ackmsg.
This messages contains all headers of the \lsadvs received.
The node then compares the \lsadvs received with the ones in its own link state database \lsdb; 
it stores those \lsadv (in the variable \lsas) that are newer than the one in the {\lsdb} (or non-existent in \lsdb);
these are the \lsadv that need to be handled.
In case there is none (Line \ref{upd_det_7}),  no action is required and the protocol returns to the main process {\OSPF} in Line~\ref{upd_det_8}.
If there are \lsadvs (Line \ref{upd_det_9}), the node installs these newly discovered \lsadvs in its own database (Line \ref{upd_det_10}), 
removes old data (\lsadvs that are dominated by others) from the data structure, and informs its neighbours about this new information 
by sending out an \lsumsg (Line~\ref{upd_det_13}).

After the updates are performed the node {\ip} checks whether there are entries in the request list that need its attention. 
In case there are some (Line \ref{upd_det_15}), 
the process returns to {\OSPF} to await a new incoming \lsumsg that address some of these \lsadvs.
Otherwise the list is empty (Line  \ref{upd_det_17}). 
That means that the node requires nothing more from its neighbour; 
it sets the status to \str{Full} (Line \ref{upd_det_18}), updates its own data structures (Lines  \ref{upd_det_19} to  \ref{upd_det_21}), 
and sends a new update to the  neighbours interested in this information (Line  \ref{upd_det_22}). 

The process \keyw{ACK} handles incoming Link State Acknowledgement ({\sc Lsa}) messages.

  \algsetup{linenodelimiter=.,linenosize=\tiny}
  \begin{algorithm}[H]
    {
     \caption{Handling \ackmsgs}
      \label{pro:detailed_ack}
      \begin{algorithmic}[1]
\DEFPROCESS{\ACK}{\lsahdrs\comma\sip\comma\ip\comma\nbrs\comma\lsdb\comma\hellot}
	\IF[the sender {\sip} is unknown]{$\neg\nbrExist\nbrs\sip$}																\label{ack_det_1}
		\ospfL{\ip}{\nbrs}{\lsdb}{\hellot}																									\label{ack_det_2}
	\ELSIF[{\sip} is a known neighbour]{$\nbrExist{\nbrs}{\sip}$}															\label{ack_det_3}
		\UPD{\nbrs := \cleanRXMTS\nbrs\sip\lsahdrs}																			\label{ack_det_4}
		\ospfL{\ip}{\nbrs}{\lsdb}{\hellot}																									\label{ack_det_5}
	\ENDIFii

		\end{algorithmic}
    }
  \end{algorithm}

As in most of the other processes, the message is dropped in case the node {\ip} does not know the originator and sender of the \ackmsg 
(Lines \ref{ack_det_1} and \ref{ack_det_2}). In case the sender in known (Line \ref{ack_det_3}) the only action that remains is
to remove {\sip} from the  link state transmission list, which is done by $\fncleanRXMTS$ in Line \ref{ack_det_4}.

\subsubsection{Message Queues}

Processes {\QMSG} and {\QSND} are the same as those in the simple model, see Processes~\ref{pro:simple_message_queue} and 
\ref{pro:simple_sending_queue}.

\newpage
\section{The Uppaal Model\label{app:modelIII}}

In this section we provide a description of our Uppaal model that was presented in \autoref{sec:modelIII}. 
The main component of the Uppaal model is a timed automaton named \keyw{ospf}. 
This automaton corresponds to the process {\OSPF} of \autoref{sec:modelII}, including all of its subprocesses ({\HELLO}, {\DBD}, etc.).
The outgoing messages  queue (Process {\QSND}), however, has its own corresponding automaton \keyw{qsnd}.
The purpose of creating a separate automaton for \keyw{qsnd} is so that the sending of a message does not block a node performing other actions while it waits for another node to be ready for synchronisation. 
Each automaton \keyw{ospf} has an associated automaton \keyw{qsnd} that handles its outgoing messages.
As a slight optimisation we are able to avoid another automaton for the input queue. 
A network of $n$ nodes is modelled by $n$ copies of the two automata; each tagged with a unique identifier. 
We use the numbers $1$ to $n$ as identifiers, we avoid the use of $0$ as it is the default initialisation of integers in \uppaal.

The Uppaal file is split into six sections: 
\textbf{(i)} The first section declares global variables, data structures and defines the topology of the network being modelled. 
\textbf{(ii)} The second defines and declares what is required locally for each automaton \keyw{ospf}; this includes function definitions and data structure declarations. 
\textbf{(iii)} A short part that defines and declares what is required for each automaton \keyw{qsnd}. 
\textbf{(iv)} A section for declaring and initialising multiple (independent) instances of all automata involved. 
\textbf{(v)} The declaration of the automaton \keyw{ospf}, using functions and data structures defined in Part (ii). 
\textbf{(vi)} The automaton for the sending queue, using data structures and functions from Part (iii).

\subsection{Global Declarations}

This section, copied verbatim from the \uppaal model,
declares the global data structure. This includes 
constants (\eg the size of the network),
type definitions (\eg message types), 
the node's own data structure (\eg the neighbour structure), and 
other concepts such as \lsadv headers and \lsadvs.
It also defines a connectivity matrix to define the topology of the network, as well as 
channels for sending messages.


\begin{uppaalcode}[basicstyle=\footnotesize,backgroundcolor=\color{backgroundColour},numbers=left]
// GLOBAL DECLARATIONS
// -------------------
// CONSTANTS
const int N=3+1;          // number of nodes +1
const int M=10;           // max sequence number +1
const int time_sending=1; // minimum time taken to send a message
const int time_spread=0;  // max time to send message is time\_sending + time\_spread
const int hellointvl=10;  // time between sending hello messages 
const int rtdeadcount=5;  // number of hellos used to decide neighbour liveness
const int start_interval = 10; // max wait time before a node initialises the
                              // ospf automaton. All routers will be booted-up
                              // nondeterministically within the start\_interval
const int age_bound = 2 * N;   // upper bound for age of an LSA. 

//TYPEDEFs
typedef int[1,N-1] IP;    // range of IP addresses in the given topology

typedef int[0,4] MSGTYPE; // definition of message types
    const MSGTYPE NONE=0;
    const MSGTYPE HELLO=1;
    const MSGTYPE DBD=2;
    const MSGTYPE REQ=3;
    const MSGTYPE UPD=4;

typedef struct // neighbour structure contains a bounded integer (inactivity timer) 
{ 
  int[0,rtdeadcount] inactivity_timer;     
} NBR;
typedef struct // LSA header contains IP of the originating router and LSA age.
{                           
  int[0,N-1] ip;
  int[0,age_bound+1] age; // age acts as a wrap-around sequence number
} LSAHDR;


typedef struct // LSA is identical to LSA header since the extra info is unused.
{              // We distinguish them as we wish to model the DBD exchange procedure
  int[0,N-1] ip;     
  int[0,age_bound+1] age;
} LSA;

typedef struct {   // definition of messages
  MSGTYPE msgtype; // the type of message
  int[0,N-1] sip;  // sender IP address, used for all message types
  bool ips[N];     // discovered neighbours, used for HELLO
  LSAHDR hdrs[N];  // used for DBD and REQ messages
  LSA lsas[N];     // used for UPD messages
} MSG;

// meta variables for copying messages. Sender copies from xlocal to xglobal
// receiver copies from xglobal to xlocal.
meta MSG msgglobal;          // this variable is for the message contents
meta bool destsglobal[N];    // this contains the intended destinations of a message

// CHANNELS
urgent chan imsg[N];                 // internal communication
broadcast chan bcast[N], gcast[N];   // channels for broadcasting and groupcasting
chan ucast[N][N];                    // channel for sending a unicast
urgent broadcast chan tau[N];        // used to prioritise internal transitions

bool send_idle[N]={1,1,1,1}; // indicates if a qsnd automaton is busy sending
                             // or else is ready to receive a message from ospf

const bool topology[N][N]={    // topology described in terms of connectivity
  {0,0,0,0},
  {0,0,1,0},
  {0,1,0,1},
  {0,0,1,0}
};

bool isconnected(IP i, IP j){
  return(topology[i][j]==1);
}
\end{uppaalcode}

\subsection{Template \texttt{\textbf{ospf}}}

The declaration of the automaton \keyw{ospf} is presented below. 
It first declares the data structures that are used (neighbour structures, LSDB, input queue, output queue, etc.). 
Afterwards it defines functions  used by \keyw{ospf}  for manipulating these data structures.

\begin{uppaalcode}[basicstyle=\footnotesize,backgroundcolor=\color{backgroundColour},numbers=left]
//  ospf DECLARATIONS
// ----------------------
// local data structure of ospf automaton
NBR nbrs[N];       // array of neighbours that have been discovered
LSA lsdb[N];       // array of LSAs (which make up the LSDB)
bool upd_required; // if a neighbouring router has died we must update the LSDB
int[1,age_bound+1] age = 1; // age of the last self-originated LSA

// internal datastructure for Uppaal
clock local_time; //local time (used to trigger hello messages)

// in-queue of messages (received from other nodes)
MSG msg_inqueue[M];     // last M local copies of incoming messages, 
int buffersize_inqueue; // number of messages currently in the in queue

// out-queue of messages to be sent
MSG msg_outqueue[M];        // Last M local copies of outgoing messages
bool dests_outqueue[M][N];  // boolean matrix of destinations for outgoing messages
int buffersize_outqueue;    // number of messages currently in the out-queue

// OUT QUEUE FUNCTIONS
// -------------------
// add message (and its destinations) to the out queue
// doesn't check if there is space in the queue; cause overflow
void addmsg_outqueue(MSG msg, bool dests[N]){
  dests_outqueue[buffersize_outqueue]=dests;
  msg_outqueue[buffersize_outqueue]=msg;
  buffersize_outqueue++;
}

// returns type of next message to be sent. If NONE, the out queue is empty
MSGTYPE nextmsg_outqueue(){
  return msg_outqueue[0].msgtype;
}

// delete message from the out-queue and shift all other messages forward one place.
void deletemsg_outqueue(){
  MSG empty;
  bool emptyd[N];
  for(i: int[1,M-1]){               // move all messages forward one position
    msg_outqueue[i-1]=msg_outqueue[i];
    dests_outqueue[i-1]=dests_outqueue[i];
  }
  msg_outqueue[M-1]=empty;          // write empty message end of the queue 
  dests_outqueue[M-1]=emptyd;
  buffersize_outqueue--;
}

// IN QUEUE FUNCTIONS
// ------------------
// add a message to the in queue
// no check for space, may cause overflow
void addmsg_inqueue(MSG msg){
  msg_inqueue[buffersize_inqueue]=msg;
  buffersize_inqueue++;
}
// returns type of next message, if it returns NONE, we know the in queue is empty
MSGTYPE nextmsg_inqueue(){
  return msg_inqueue[0].msgtype;
}

// delete message from the in-queue and shift all other messages forward one place.
void deletemsg_inqueue(){
  MSG empty;
  for(i: int[1,M-1]){               // move all messages by one position
    msg_inqueue[i-1]=msg_inqueue[i];
  }
  msg_inqueue[M-1]=empty;           // write an empty message to end of the queue 
  buffersize_inqueue--;
}

// MISC
// ----
void initialise(){ // when a node boots-up it will its initial LSA to the LSDB
  lsdb[ip].ip = ip;
  lsdb[ip].age = 1;
}


// OSPF SPECIFIC
// -------------
// create new LSA for own IP
LSA newLSA(){
  LSA lsa;
  lsa.ip = ip;
  age==age_bound?age=1:age++; // wrap around age if exceeds age\_bound
  lsa.age = age;              
  return lsa;
}

// returns true if age1 is newer than age2 (handles wrap around)
bool newer_age(int age1, int age2){
  if (age1==0) { return false; } 
  else if (age2==0) { return true; }
  if ((age2 > age1 && 2*(age2 - age1) < age_bound) // wrap around handling
     || 
     (age1 > age2 && 2*(age1 - age2) > age_bound))
      return false;
  return true;
}

void install_lsa(LSA lsa){ // install an LSA in LSDB if newer than current copy
  if(newer_age(lsa.age,lsdb[lsa.ip].age))
      lsdb[lsa.ip]=lsa;
}

void install_lsas(LSA lsas[N]){ // install multiple LSAs into the LSDB if newer
  for(i : IP){
    if(newer_age(lsas[i].age,lsdb[i].age))
      lsdb[i]=lsas[i];
  }
}
void reduce_lsas_msglocal(){ // check array of LSAs against LSDB, return newer LSAs
  LSA fresh;
  for(i : IP)
    if(newer_age(lsdb[i].age, msg_inqueue[0].lsas[i].age))
      msg_inqueue[0].lsas[i] = fresh;
}

bool lsas_is_empty(){ // after reduce\_lsas\_msglocal, check if any LSAs remain
  LSA fresh[N];
  return(fresh==msg_inqueue[0].lsas);
}

bool nbrsExist(){ // check if sender of current message is active
  return(nbrs[msg_inqueue[0].sip].inactivity_timer>0);
}

void reduce_lifetime_nbrs(){ // after a hello is sent, reduce all inactivity timers
  for (i : IP){
    if(nbrs[i].inactivity_timer > 0){
      nbrs[i].inactivity_timer--;
      if(nbrs[i].inactivity_timer == 0){ // if timer is zero, set upd\_required flag
        upd_required=true;  
       }
    }
  }
}


// CREATE MESSAGES
// ---------------
// create hello message and add it to out queue
void generate_send_hello(){ 
  MSG msg;
  bool dests[N];
  msg.msgtype=HELLO;
  msg.sip=ip;
  for(i:IP)
    if(nbrs[i].inactivity_timer!=0)
     msg.ips[i]=true;
  addmsg_outqueue(msg,dests);
}

// create dbd message, add to out queue
void generate_send_dbd(IP dip){ 
  MSG msg;
  bool dests[N];
  msg.msgtype=DBD;
  msg.sip=ip;
  for(i:IP){
    msg.hdrs[i].ip = lsdb[i].ip;
    msg.hdrs[i].age = lsdb[i].age;
  }
  dests[dip]=true;
  addmsg_outqueue(msg,dests);
}

void generate_send_upd_single(LSA lsa){//create single-LSA upd msg, put in out queue
  MSG msg;
  bool dests[N];
  msg.msgtype=UPD;
  msg.sip=ip;
  msg.lsas[lsa.ip] = lsa;
  for(i : IP)
    dests[i] = (nbrs[i].inactivity_timer!=0);
  addmsg_outqueue(msg,dests);
}

void generate_send_upd(LSA lsas[N]){ // create (multi-LSA) upd msg, add to out queue
  MSG msg;
  bool dests[N];
  msg.msgtype=UPD;
  msg.sip=ip;
  msg.lsas = lsas;
  for(i : IP)
    dests[i] = (nbrs[i].inactivity_timer!=0);
  addmsg_outqueue(msg,dests);
}

// create and process an upd message in response to a request message
void generate_send_upd_requested(IP dip){
  MSG msg;
  bool dests[N];                          
  msg.msgtype=UPD;
  msg.sip=ip;
  for(i:IP)
    if(msg_inqueue[0].hdrs[i].ip == i && newer_age(lsdb[i].age,msg_inqueue[0].hdrs[i].age))
      msg.lsas[i] = lsdb[i];
  dests[dip]=true;
  addmsg_outqueue(msg,dests);
}

// after receiving a dbd message create a req message to request required LSAs
void generate_send_req(IP dip){
  MSG msg; 
  bool dests[N];
  LSAHDR fresh[N];
  msg.msgtype=REQ;
  msg.sip=ip;
  for(i : IP)
    if(newer_age(msg_inqueue[0].hdrs[i].age, lsdb[i].age))
      msg.hdrs[i] = msg_inqueue[0].hdrs[i];

  if(msg.hdrs!=fresh){   // only send if request is non-empty
    dests[dip]=true;
    addmsg_outqueue(msg,dests);
  }
}
\end{uppaalcode}

\subsection{Template \texttt{\textbf{qsnd}}}

The declaration of the automaton \keyw{qsnd} is much shorter, but structured in the same way as \keyw{ospf}.
\begin{uppaalcode}[basicstyle=\footnotesize,backgroundcolor=\color{backgroundColour},numbers=left]
// qsnd DECLARATIONS
// -----------------
MSG msglocal;
bool dests[N];
clock clk;

int getDip(){ // retrieve destination of message to be sent
  for(i : IP)
    if(dests[i]) return i;
  return 0;
}

void clean_dests(){ // after sending clear the destination from the global store
  bool empty[N];
  destsglobal=empty;
}
\end{uppaalcode}

\subsection{System Declarations}

The system declarations are used to instantiate the required number automata. In our example, the instantiation of a network consists of three nodes 
each running both \keyw{ospf} and \keyw{qsnd}.

\begin{uppaalcode}[basicstyle=\footnotesize,backgroundcolor=\color{backgroundColour},numbers=left]
// system DECLARATIONS
// -------------------
// Place template instantiations here.
a1ospf=ospf(1); // instantiate a ospf automaton called a1ospf
a1send=qsnd(1); // instantiate a qsnd automaton called a1send

a2ospf=ospf(2);
a2send=qsnd(2);

a3ospf=ospf(3);
a3send=qsnd(3);

system a1ospf,a1send,a2ospf,a2send,a3ospf,a3send;
\end{uppaalcode}

\subsection{The Automaton \texttt{\textbf{ospf}}}

\begin{figure}
\centering
\includegraphics[width=\textwidth]{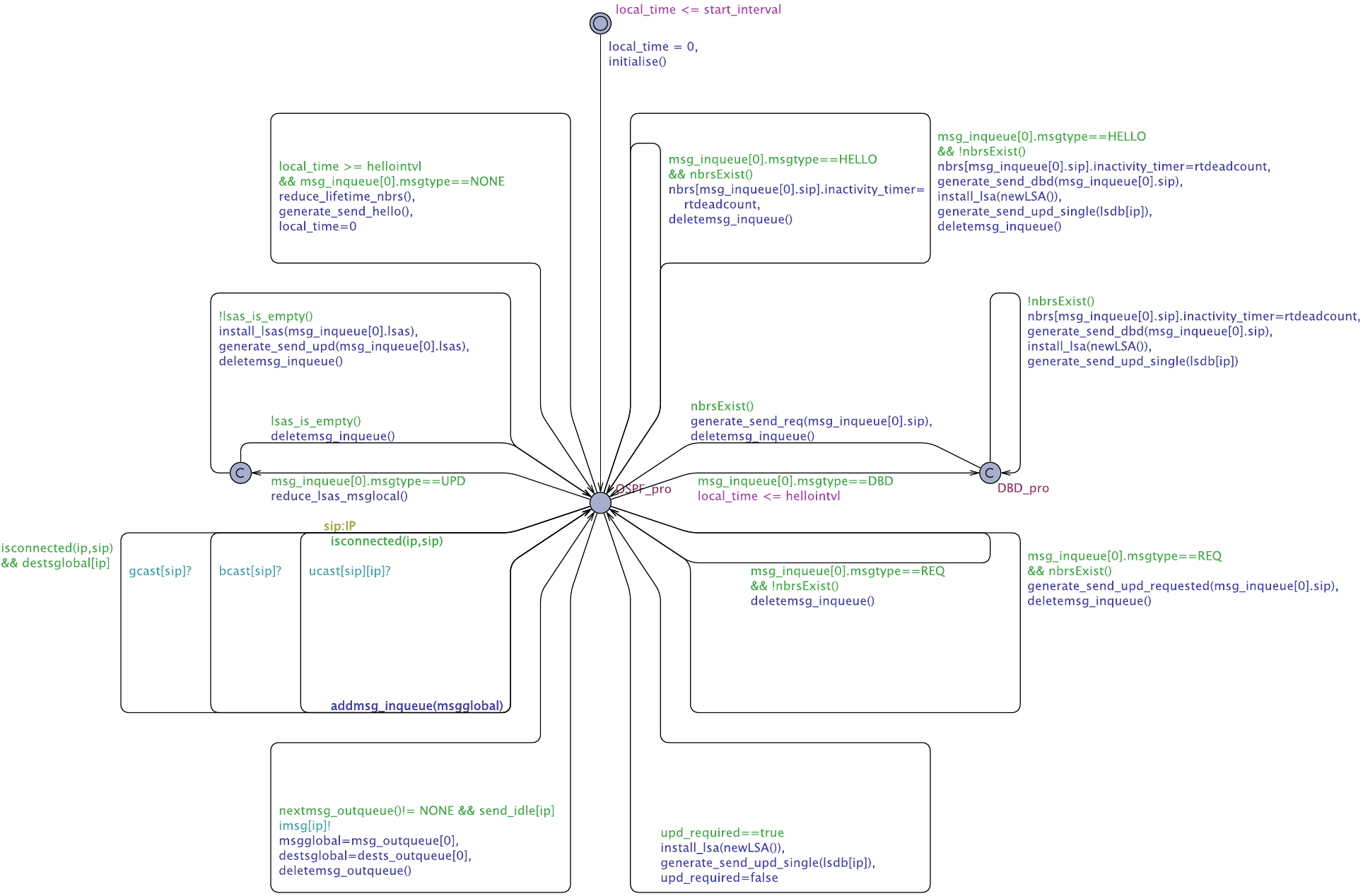}
\caption{The timed automaton \keyw{ospf}\label{fig:appuppaal1}}
\end{figure}

The automaton \keyw{ospf} is presented in \autoref{fig:appuppaal1}. The initial state is represented as the node at the top of the image with only a single, downward, outgoing transition. The automaton leaves this node after a non-deterministic amount of time, bounded by \keyw{start\_interval}.

The node at the centre represents a state where the automaton is ready to handle any message or action required.
There are eight loops consisting of multiple transitions each, that surround the centre.
We address them in a clockwise fashion beginning from one o'clock. 
The first three loops handle {\sc Hello}, {\sc Dbd} and \lsrmsgs (here called \keyw{REQ}).
It is easy to see how the synchronisation and updates performed correspond to the actions of our simple model, explained in \autoref{app:modelII}.
 The fourth loop is executed when a neighbouring node dies and therefore a new LSA needs to be generated. The fifth pushes a message from the output queue to the relevant automaton \keyw{qsnd}. The sixth deals with receiving a message and placing it into the input queue. The seventh handles a \lsumsg (here called \keyw{UPD}). The eighth and final transition loop handles the generation of a \hellomsg at the correct time.

\subsection{The Automaton \texttt{\textbf{qsnd}}}
\begin{figure}[h]
\centering
\includegraphics[width=.82\textwidth]{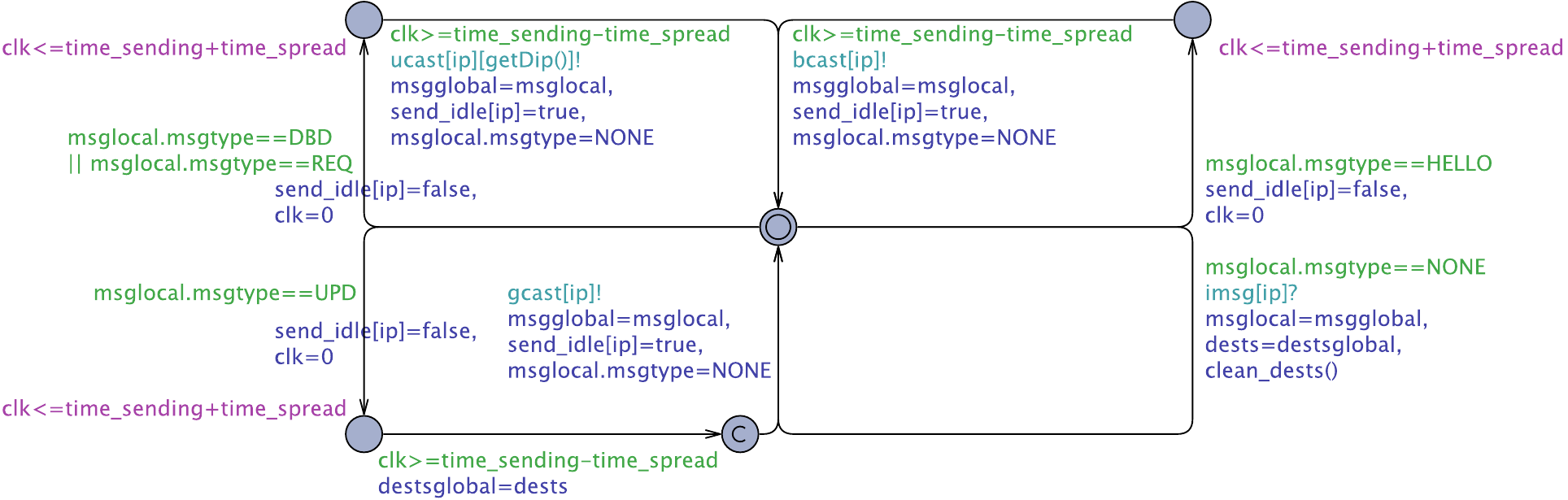}
\caption{The timed automaton \keyw{qsnd}\label{fig:appuppaal2}}
\end{figure}

The \keyw{qsnd} automaton is presented in \autoref{fig:appuppaal2}. Its purpose is to hold messages until other automata are ready to receive them. Its initial state is at the centre. The top right quadrant broadcasts {\HELLO} messages. The top left quadrant unicasts {\DBD} and {\REQ} messages. The bottom left quadrant groupcasts {\UPDp} messages. The bottom right quadrant receives a message from the corresponding \keyw{ospf} automaton.

\end{document}